\documentclass[fleqn,usenatbib]{mnras}

\usepackage[T1]{fontenc}

\DeclareRobustCommand{\VAN}[3]{#2}
\let\VANthebibliography\thebibliography
\def\thebibliography{\DeclareRobustCommand{\VAN}[3]{##3}\VANthebibliography}

\usepackage{graphicx}	
\usepackage{amsmath}	
\usepackage{amssymb}	
\usepackage{newtxtext,newtxmath}

\newcommand{\hi}{\text{H\,\sc{i}}}
\newcommand{\gad}{{\sc Gadget-3}}
\newcommand{\gizmo}{{\sc Gizmo}}

\newcommand{\simba}{{\sc Simba}}
\newcommand{\martini}{{\sc Martini}}
\newcommand{\eagle}{{\sc Eagle}}
\newcommand{\caesar}{{\sc Caesar}}

\newcommand{\hmpc}{\,h^{-1}{\rm Mpc}}

\title[ASymba: HI global profile asymmetries in Simba]{ASymba: HI global profile asymmetries in the \simba\ simulation}

\author[]{M. Glowacki$^{1,2,3}$\thanks{E-mail: marcin.glowacki@curtin.edu.au}, N. Deg$^{4}$\thanks{E-mail: nathan.deg@queensu.ca}, S.-L. Blyth$^{5}$, N. Hank$^{6}$, R. Davé$^{7,3}$, E. Elson$^{3}$, K. Spekkens$^{8}$
\\
$^{1}$International Centre for Radio Astronomy Research, Curtin University, Bentley, WA 6102, Australia\\
$^{2}$Inter-University Institute for Data Intensive Astronomy (IDIA), South Africa\\
$^{3}$Department of Physics and Astronomy, University of the Western Cape, Robert Sobukwe Road, Bellville 7535, South Africa\\
$^{4}$Department of Physics, Engineering Physics, and Astronomy, Queen's University, Kingston, ON, K7L 3N6, Canada\\
$^{5}$Department of Astronomy, University of Cape Town, Private Bag X3, Rondebosch 7701, South Africa\\
$^{6}$Kapteyn Astronomical Institute, University of Groningen, Landleven 12, 9747 AD Groningen, the Netherlands\\
$^{7}$Institute for Astronomy, University of Edinburgh, Royal Observatory, Blackford Hill, Edinburgh, EH9 3HJ, UK\\
$^{8}$Department of Physics and Space Science, Royal Military College of Canada, P.O.\ Box 17000, Station Forces Kingston ON K7K~7B4, Canada\\
}

\date{Accepted 2022 September 14. Received 2022 September 13; in original form 2022 April 29}

\pubyear{2022}

\begin{document}
\label{firstpage}
\pagerange{\pageref{firstpage}--\pageref{lastpage}}
\maketitle

\begin{abstract}
Asymmetry in the spatially integrated, 1D \hi\ global profiles of galaxies can inform us on both internal (e.g. outflows) and external (e.g. mergers, tidal interactions, ram pressure stripping) processes that shape galaxy evolution. Understanding which of these primarily drive \hi\ profile asymmetry is of particular interest. In the lead-up to SKA pathfinder and SKA \hi\ emission surveys, hydrodynamical simulations have proved to be a useful resource for such studies. Here we present the methodology behind, as well as first results, of ASymba: Asymmetries in \hi\ of \simba\ galaxies, the first time this simulation suite has been used for this type of study.
We generate mock observations of the \hi\ content of these galaxies and calculate the profile asymmetries using three different methods. We find that $M_{\rm HI}$ has the strongest correlation with all asymmetry measures, with weaker correlations also found with the number of mergers a galaxy has undergone, and gas and galaxy rotation. We also find good agreement with the xGASS sample, in that galaxies with highly asymmetric profiles tend to have lower \hi\ gas fractions than galaxies with symmetric profiles, and additionally find the same holds in sSFR parameter space. For low \hi\ mass galaxies, it is difficult to distinguish between asymmetric and symmetric galaxies, but this becomes achievable in the high \hi\ mass population. These results showcase the potential of ASymba and provide the groundwork for further studies, including comparison to upcoming large \hi\ emission surveys.
\end{abstract}




\section{Introduction}

The neutral hydrogen (\hi) gas in galaxy disks typically extends further than the stellar distribution and is more susceptible than the stars to disturbance from environmental processes. In dense environments, processes such as ram-pressure stripping \citep{GunnGott1972}, tidal interactions, and galaxy-galaxy interactions and mergers can result in asymmetric morphologies in both the stellar and gas components of galaxies \cite[e.g.][]{Deg2020}. Accretion \citep{Sancisi2008} and outflows \citep{Fraternali2017} can also lead to asymmetries in the \hi\ distributions. These asymmetries can be directly observed in both the spatial (2D) and spectral (1D global profile) \hi\ distributions. 

Early studies \citep{Peterson1974, Tifft1988, RichterSancisi1994, Haynes1998, Matthews1998} focused on using \hi\ global profiles to measure \hi\ asymmetries due to the larger samples of single-dish data available compared to imaging data. This has continued to more recent studies which have used even larger samples from single-dish \hi\ surveys such as HIPASS \citep{Meyer2004}, ALFALFA \citep{Haynes2018} and xGASS \citep{Catinella2018}. At the same time, the increase in \hi\ imaging surveys has recently enabled the measurement of 2D asymmetries, more in line with optical techniques \citep{Holwerda2011,Lelli2014,Giese2016}. \hi\ profile asymmetries have been found in isolated galaxy samples \citep{Espada2011}, to be relatively common in the field \citep{RichterSancisi1994,Matthews1998}, enhanced in close merger-pairs compared to isolated galaxies \citep{Bok2019}, and to depend on local environmental density \citep{Reynolds2020}, implying a range of different processes at work to create them. A recent study by \cite{Zuo2022} found no obvious excess in asymmetry of their merger galaxy sample compared to a sample of non-merging galaxies, underlining the variety of processes that must be giving rise to profile asymmetries. Therefore, studying \hi\ asymmetries in different galaxy samples in different environments should help to shed light on the physical processes driving galaxy evolution. 


The SKA pathfinder telescopes will present us with deeper, more sensitive \hi\ observations, in which the asymmetry of \hi\ global profiles can also be studied. Ongoing large \hi\ surveys such as the Widefield ASKAP L-band Legacy All-sky Blind surveY \citep[WALLABY;][]{Koribalski2020} and the Deep Investigation of Neutral Gas Origins \cite[DINGO;][]{Meyer2009} on the Australian SKA Pathfinder telescope \cite[ASKAP;][]{Deboer2009}, as well as the MeerKAT International GHz Tiered Extragalactic Exploration \hi\ survey \cite[MIGHTEE-HI;][]{Jarvis2017,Maddox2021} and Looking At the Distant Universe with the MeerKAT Array \cite[LADUMA;][]{Blyth2016} survey on the MeerKAT radio  telescope \citep{Jonas2016} will probe large cosmic volumes over a range of redshifts, observing many thousands of galaxy \hi\ global profiles enabling redshift evolution studies of asymmetry. The majority of detections, particularly at higher redshifts, will be spatially unresolved although the \hi\ spectra will be available. Therefore, it is important to consider what we can use with the global \hi\ profiles alone.


These surveys are in the preliminary stages and are years away from completion. 
However, cosmological simulations can provide insights into the underlying physical processes leading to the observed properties of galaxies. Unlike existing observational samples we can greatly extend sample sizes of spatially resolved galaxies (in the thousands), which can hence overcome any biases towards gas-rich observations. The ability to easily access galaxy properties that require multi-wavelength studies in reality, and accurate environmental information, enables us both to compare to existing, less sensitive studies, and to make predictions for upcoming surveys. 

There are many different processes that can disturb the \hi\ distribution of a galaxy, and determining which of these processes tend to dominate or drive asymmetry is of particular interest.  Recently \cite{Watts2020} generated mock profiles from the IllustrisTNG simulation \citep{Weinberger2017,Pillepich2018} and found that TNG100 galaxies typically have \hi\ profiles that are not fully symmetric, and that satellite galaxies are more asymmetric than centrals. The effect is primarily driven by the satellite population within a virial radius of massive haloes, typical of medium and large galaxy groups. This demonstrates the importance of deeper \hi\ emission survey studies with SKA pathfinder telescopes already underway. Another key finding of \cite{Watts2020} is that asymmetries are not driven solely by environment, but also multiple physical processes. 

\citet{Manuwal2021} examined profile asymmetries in the \eagle\ simulation \citep{Crain2015,Schaye2015,McAlpine2016}. They used a variety of methods to quantify the asymmetries and found, like \cite{Watts2020}, that satellite galaxies tend to be more asymmetric than central galaxies.  This difference was attributed to ram-pressure and tidal stripping and not to satellite-satellite interactions.  They did not find a significant difference in asymmetries as a function of stellar mass, but rather that, for a given stellar mass, galaxies with symmetric \hi\ profiles are more gas rich and show a different trend in specific star formation rate vs stellar mass compared to asymmetric galaxies. For centrals, they also found that asymmetric profiles tend to be found in younger, less-relaxed haloes. And, for a given halo mass, asymmetric galaxies host a larger number of subhaloes and show larger degrees of gas accretion as well as outflows.

\citet{Bilimogga2022} also recently used mock galaxies from \eagle\ to investigate how measured \hi\ asymmetries depend on various observational constraints such as resolution, signal-to-noise and the column density of the observations. They determined limits for these variables which result in reliable measurements for both 2D and global profile asymmetries.

In this work we present the first results of 1D asymmetry studies in a different sample of simulated galaxies, from the \simba\ simulation suite \citep{Dave2019}. \simba\ has been shown to replicate observations of cold gas in galaxies well \citep{Dave2020}. One example of this is the favourable comparison between \simba\ and ALFALFA \citep{Haynes2018}, relative to EAGLE and IllustrisTNG, for the \hi\ mass function \cite[see fig.~3 of][]{Dave2020}. This paper marks the first of many planned in this new project, henceforth named ASymba (ASymmetries in \hi\ of \simba\ galaxies). In Section~\ref{Sec:2} we introduce our sample, our method of generating mock \hi\ cubes, and their corresponding profiles. Section~\ref{Sec:Profiles} contains our definitions of different velocity profile asymmetries. In Section~\ref{Sec:GenTrends} we then discuss general trends found for our \simba\ galaxies, and in Section~\ref{Sec:Watts} we compare our sample to an observational study of \hi\ asymmetry. Finally in Section~\ref{Sec:Conclusions} we summarise our findings and discuss upcoming works for ASymba. 

\section{Simulations, \hi\ cubes and spectra}\label{Sec:2}
\subsection{SIMBA}\label{Sec:simba}

The \simba\ simulation suite \citep{Dave2019} is a cosmological hydrodynamic simulation based upon the \gizmo\ code~\citep{Hopkins2015}, which itself is an offshoot of \gad~\citep{Springel2005}. \gizmo\ uses a meshless finite mass (MFM) hydrodynamics solver that is shown to have advantageous features over Smoothed Particle Hydrodynamics and Cartesian mesh codes, such as the ability to evolve equilibrium disks for many dynamical times without numerical fragmentation~\citep{Hopkins2015}. We direct the reader to \cite{Dave2019} for further details on the \simba\ simulation suite. 

\begin{figure}
\centering
    \includegraphics[width=0.99\linewidth]{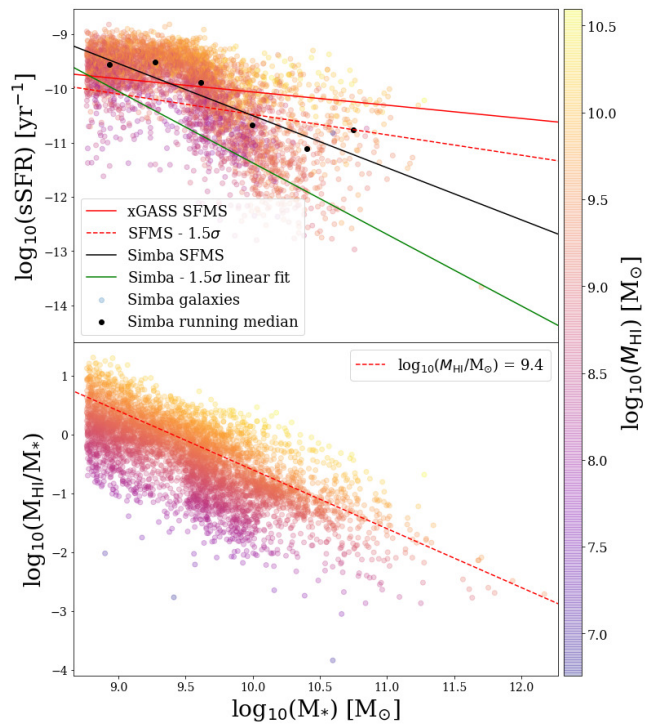} 
\caption{Scaling relations and property distributions of our \simba\ sample. Top: sSFR as a function of stellar mass. We give the xGASS SFMS and the SFMS-1.5$\sigma$ relations (red solid and dashed lines) used to define a subsample in Section~\ref{Sec:Watts}. We also give the running median of the \simba\ SFMS, its linear fit (black solid line), and SFMS-1.5$\sigma$ relation (solid green line).} Bottom: The \hi\ mass fraction scaling relation. The red line in this panel corresponds to a constant log($M_{\rm HI}$/$\mathrm{M}_\odot$)~=~9.4, as shown in fig.~2 of \citet{Watts2021} and considered in mass cuts in Section~\ref{Sec:Watts}.
  \label{Fig:SimbaSample}
\end{figure}

For this study we adopt the $(50\hmpc)^3$ periodic volume, with $512^3$ dark matter particles and $512^3$ gas elements with full \simba\ feedback mechanisms implemented; the impact of variants on feedback models on asymmetry measures that are only available with this sized snapshot in \simba\ will be explored in separate studies. The assumed cosmology is concordant with~\citet{Planck2016}: \(\Omega_M = 0.3\), \(\Omega_{\Lambda} = 0.7\), \(\Omega_{b} = 0.048\), \(H_0 = 68\) km s\(^{-1}\) Mpch\(^{-1}\), \(\sigma_8 = 0.82\), \(n_s = 0.97\).  This yields a mass resolution of \(9.6\times 10^7 M_{\odot}\) for dark matter particles and \(1.82\times 10^7 M_{\odot}\) for gas elements. Adaptive gravitational\, comoving softening length is employed with a minimum \(\epsilon_{\rm{min}} = 0.5h^{-1}\)c\,kpc.

Only the redshift $z$~=~0 snapshot is considered in this analysis. Galaxies were identified via a 6-D friends of friends (FOF) algorithm, and their corresponding halos identified via a 3-D FOF algorithm. Galaxies and haloes were cross-matched and their properties computed using \caesar\footnote{\tt https://caesar.readthedocs.io/en/latest/}, a particle-based extension to {\sc yt} \citep{Turk2011}. \hi\ is associated with each galaxy by summing the \hi\ content gravitationally bound to that galaxy, from all gas particles within its respective halo. We note that the \hi\ fraction in each gas particle is computed in \simba, accounting for self-shielding on the fly, based on the prescription in \cite{Rahmati2013}, and includes photoionisation from a spatially uniform ionising background given by \cite{Haardt2012}.

We set a minimum stellar mass limit of $M_{*}$~$>$~5.8$\times$10$^{8}$\,$M_{\odot}$ \cite[the galaxy stellar mass resolution limit for the \simba\ simulations considered here;][]{Dave2019}, and require an \hi\ mass of $M_{\rm{\hi}}$~$>$~1$\times$10$^{7}$\,$M_{\odot}$, as we would not otherwise be able to construct an \hi\ cube for this analysis for galaxies lacking in cold neutral atomic hydrogen. We present the sSFR-$M_{*}$ and \hi\ mass fraction-$M_{*}$ scaling relation for our sample in Fig.~\ref{Fig:SimbaSample}. This plot purposefully mimics that of fig.~2 of \cite{Watts2021} by including the xGASS star forming main sequence (SFMS) relation and other lines, to better illustrate a subsample we construct and compare to the \cite{Watts2021} study of the xGASS sample in Section~\ref{Sec:Watts}. We note that structure in the figure is attributed to the seeding of black holes in \simba\ at log($M_{*}$)~$\sim$~9.5\,$\mathrm{M}_\odot$ which results in abrupt transitions in various properties; see \cite{Dave2019} for further details.

Mergers are also identified via tracking progenitors of galaxies across 46 snapshots back to $z = 1$, via finding the two galaxies with the most star particles in common with the descendant galaxy. Note that we consider all galaxies down to $M_{*}$~=~2.9~$\times$~10$^{8}$~M$_{\odot}$, half the nominal $M_{*}$ resolution limit which is the mass limit down to which Caesar identifies galaxies, so the smallest galaxies may not have a fully representative merger count.

\subsection{Spectral line cubes}\label{Sec:cubes}

We generated \hi\ cubes via \martini\footnote{\url{https://github.com/kyleaoman/martini}, version 1.5}, in a similar manner to that in \cite{Glowacki2021}. \martini\ is a package for creating synthetic resolved \hi\ line observations -- aka data cubes -- of smoothed particle hydrodynamical simulations of galaxies \citep{Oman2019}. It is ideal as it allows for realistic mock observations with all the aforementioned specifications implemented. \martini\ achieves this by taking the input \simba\ snapshot file and accompanying \caesar\ catalogue, which contains the galaxy and host halo properties including \hi\ fraction values. 

With \martini\, there are a few approaches possible to construct \hi\ spectral line cubes. Here we outline our process for clarity:

\begin{itemize}
    \item We make a separate cube for every target galaxy in our sample and
    we opt to mimic the fact that in real observations spectral line cubes will inevitably include some contributions from nearby galaxies and the corresponding outer halo in addition to the contribution from the individual target galaxy. Therefore, all cubes we create include all the \hi\ flux in the specific sub-volume of the simulation, irrespective of any satellites that may be present. We also do not exclude mergers from our sample. We later apply a SoFiA~2 run on each cube to isolate sources as done in real observations; see Section~\ref{Sec:Profiles} for further details. 
    \item A dynamic aperture is used, in that a larger box size and number of spectral line channels is used for more (\hi) massive galaxies in our sample, and a smaller aperture as mass decreases. Essentially, the aperture was adjusted and cubes remade to match the extent of each galaxy determined from its corresponding \hi\ moment maps (intensity and velocity) in an iterative manner. In the line-of-sight direction, the aperture size in \martini\ is set to 100~kpc.
    \item The initial spectral line cubes are constructed to mimic the `32k' spectral line mode of MeerKAT ($\sim$5.51~km\,s$^{-1}$) to aid in future comparisons with observations from the LADUMA and MIGHTEE-HI surveys.
    We note that smoothing is done for some asymmetry measures; see Section~\ref{Sec:Profiles}. Cubes are convolved with a typical radio beam of $\sim$10 arcseconds, with an assumed distance of 4~Mpc. Explicitly, BMAJ and BMIN, the major and minor axis of the radio beam, are set to 11.2 and 9.8~arcseconds, as per early L-band MeerKAT observations and data products of LADUMA (private communication). A variable size of velocity channels (50 to 200) were used for cubes, to balance both cube creation time in \martini\ and produce cubes containing all \hi\ emission associated with the galaxy. A temperature-dependent Gaussian line profile is assumed.
    \item In order to measure the maximum asymmetry from the spectral line, all galaxies were orientated to be edge-on to the observer in their resulting spectral line cube, so that no inclination correction is required. This is done through the use of the \hi\ gas angular momentum vector, and may rarely result in a missed orientation due to extended gas structures. This is done in \martini\ by considering the angular momentum of the inner 30\% of particles (by \hi\ mass) to define the plane of the disc.  Naturally, this idealised edge-on scenario is not the case with observations, but this does enable a more direct comparison of asymmetries to intrinsic galaxy properties. 
    \item To obtain measurements as close to the `true' asymmetry as possible,
    we do not simulate noise in the data cubes. This is another aspect to bear in mind in our analysis for the low \hi\ mass end of our sample, although when comparing to results from observations by \cite{Watts2021} in Section~\ref{Sec:Watts} we include the same minimum mass limits on our \simba\ sample. Despite the lack of noise added to the mock observations, it is possible that particle `shot' noise may affect the calculated profiles and thereby asymmetries.  \citet{Watts2020} examined this effect in \textsc{IllustrisTNG} and found that $\ge 500$ gas cells per galaxy were required to minimize this type of noise.  In our sample, only a handful of galaxies have $\le 500$ gas particles and the lowest number of particles in an observation is 359.  Thus, we do not expect particle `shot' noise to strongly affect our analysis.
\end{itemize}



\subsection{Mock profiles}\label{Sec:Profiles}

For `noiseless' cubes, it is relatively straightforward to calculate a mock profile.  However, as mentioned in Section~\ref{Sec:cubes} the cubes may contain emission in addition to that from the target galaxy due to gas from nearby neighbours/satellites, accretion, etc. As would be done for real observations to separate the target galaxy emission from other emission, we ran the \textsc{SoFiA 2} (HI Source Finding Application) code  \citep{Westmeier2021} on each of the mock cubes.

Running \textsc{SoFiA 2} on these mock cubes is non-trivial as the code assumes that the cubes have some noise to them.  In our case, it is not necessary to add realistic noise to the cubes, as the profiles are constructed from the noiseless cubes.  \textsc{SoFiA 2} only requires that random fluctuations are present in order to find and separate different objects in the cube itself.  The procedure we have adopted here to produce `noisy' cubes is:
\begin{enumerate}
    \item calculate the total flux and number of cells in the noiseless cubes;
    \item set a noise value $\sigma=f F_{\rm{tot}}/n_{\rm{cells}}$.  The factor $f$ determines the relative strength of the noise. In practice, we found $f=15$ provides good results in terms of separation and flux recovery;
    \item generate a noise-plus-signal cube using Gaussian random draws with width $\sigma$;
    \item run \textsc{SoFiA 2} on the noisy cube. If \textsc{SoFiA 2} finds multiple objects inside the cube, select the object with the largest total flux;
    \item apply the mask for the largest flux detection to the original noiseless cube;
    \item construct a noiseless profile using the masked noiseless cube. These noiseless profiles are used in all further analysis.
\end{enumerate}
The noise in the noise-plus-signal cubes is not beam smeared and is not tied to any observational limit.  The goal here is simply to provide \textsc{SoFiA 2} a cube with a measurable amount of noise so that it can separate out extraneous gas.

\begin{figure*}
\centering
    \includegraphics[width=0.99\linewidth]{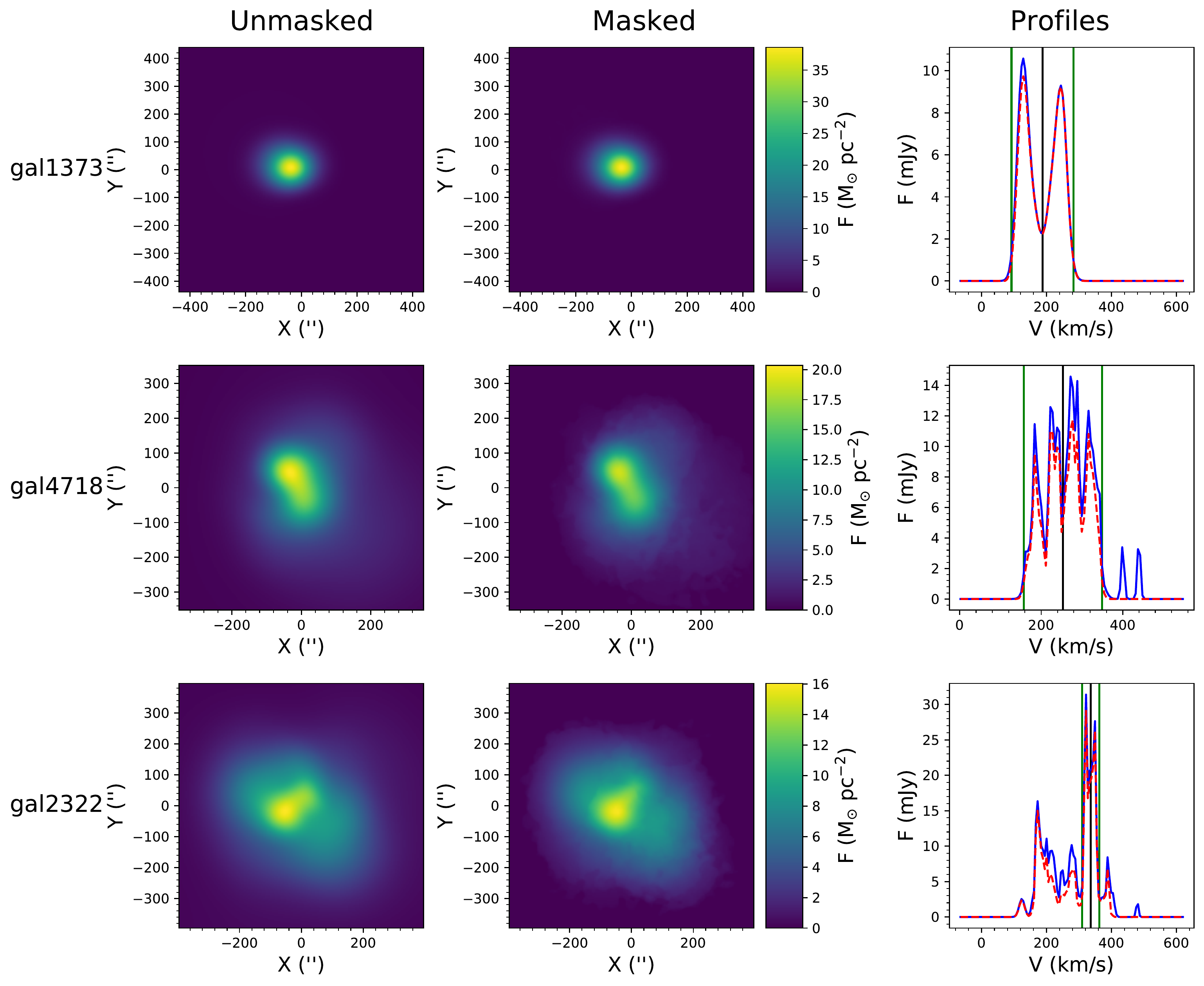} 
\caption{A sample set of galaxy maps and profiles.  The left-most column shows the \hi\ moment 0 map for the three selected galaxies and the middle column shows the same moment maps after applying the \textsc{SoFiA} mask to the cubes.  The right hand column shows the corresponding velocity profiles.  The blue solid line is the unmasked profile and the red dashed line is the masked profile.  The vertical black line is the calculated $v_{\rm{sys}}$ and the green lines show the edges of the profiles that are used in the asymmetry calculations.  For galaxy 4718, the masking process removes the extra set of flux spikes located at $v>400$ km/s.  For galaxy 2322 the edges are found around the central spike associated with the bright center in the moment maps and excludes the more diffuse gas.  This means that the asymmetry calculations will only include the central flux.}
  \label{Fig:SampleProfiles}
\end{figure*}

Figure \ref{Fig:SampleProfiles} shows the effect of this process on some sample profiles.  These particular samples have been selected to highlight different situations that may arise when characterizing the profile as well as the effect of masking. 
An examination of all three rows shows that the masking does remove some gas associated with other galaxies, while retaining most of the flux.  In particular, the masking has removed the pair of high velocity peaks in the second row.  Many of our galaxies are similar to the upper row, where the system is well behaved and the profile shows a clear double horned profile.  However, just as many galaxies show multiple peaks.  This is due to the \hi\ gas potentially being confused due to ongoing mergers/accretion and the gas not being relaxed as in the 2nd and 3rd rows in Fig.~\ref{Fig:SampleProfiles}. As noted in Sec. \ref{Sec:simba} all observations, including these three galaxies, are set to have edge-on inclinations using the angular momentum of \hi\ gas. Thus even in cases with unrelaxed gas, such as the 2nd and 3rd rows of Fig. \ref{Fig:SampleProfiles}, where the Mom0 map does not appear edge-on, the profiles will cover as many velocity channels as possible.

Given the complicated nature of many of the profiles, great care must be taken when calculating the profile edges, $v_{l}$ and $v_{h}$, and systemic velocity, $v_{\rm{sys}}$.  While we could simply set $v_{\rm{sys}}$ to the velocity of the halo, we opted to follow a more observationally motivated approach.  For this procedure we:
\begin{enumerate}
    \item generate a `smoothed' profile using a Gaussian kernel (initially 3 channels wide, but can be made wider if necessary);
    \item use the smoothed profile to estimate the slope of the profile at each velocity value. This smoothing step is necessary in order to avoid poor estimates of the profile slopes in the next step due to channel-by-channel fluctations like some of those seen in Fig. \ref{Fig:SampleProfiles}.
    \item use the location of the minimum and maximum profile slopes as a first estimate of the profile edges, $v_{l,e}$, and $v_{h,e}$;
    \item use the estimated edges to estimate the systemic velocity via $v_{\rm{sys},e}=(v_{l,e}+v_{h,e})/2$;
    \item use the smoothed profile to estimate if it has a single peak or multiple peaks;
    \item Find the location of the singular peak or find the two peaks in the ranges $v_{l,e}-v_{\rm{sys},e}$ and $v_{\rm{sys},e}-v_{h,e}$ using the unsmoothed profile;
    \item determine the measured edges, $v_{l}$ and $v_{h}$, as the points where each $F_{\rm{edge}}=0.1F_{\rm{peak}}$ for each peak independently, again using the unsmoothed profile.  This limit may occasionally select the central portion of a galaxy that has a large amount of diffuse gas around it.  An example of this is shown in the gal2322 panels in Fig. \ref{Fig:SampleProfiles}.
    \item use the measured edges to obtain the correct value for $v_{\rm{sys}}$.
\end{enumerate}

If a profile contains a very narrow peak, it is possible that the initial edge estimation will be too narrow.  Thus, if the initial width is $<15$ channels, we attempt to increase the smoothing to first 5 and then 7 channels.  If this still fails, the profile is discarded from the sample.  Similarly, if the profile contains any empty channels between $v_{l}$ and $v_{h}$, it is also discarded.  These cuts remove a total of 246 galaxies and we are left with a sample of 4264 profiles.  This is broadly comparable to the sample size studied in \citet{Manuwal2021} constructed from the EAGLE simulation and roughly 40\% of the sample size constructed from IllustrisTNG100 simulation used in \citet{Watts2020}.

\section{Profile Asymmetries}

Estimating how asymmetric an \hi\ velocity profile appears is a somewhat old question that has been approached using both quantitative methods \citep{Peterson1974,Haynes1998,Matthews1998,Reynolds2020,Deg2020,Yu2020} and visual inspection (e.g., \citealt{RichterSancisi1994}).  The first measurement of profile asymmetry is the profile lopsidedness or flux ratio, $A$ of \citet{Peterson1974}.  There are a number of different versions of this quantity. For this work we adopt
\begin{equation}\label{Eq:Lopsidedness}
A_{L}=\frac{|F_{l}-F_{h}|}{F_{l}+F_{h}}~,
\end{equation}
where 
\begin{equation}
    F_{l}=\int_{v_{l}}^{v_{\rm{sys}}} F(v) dv~,
\end{equation}
and 
\begin{equation}
    F_{h}=\int_{v_{\rm{sys}}}^{v_{h}} F(v) dv~.
\end{equation}
An advantage of Eq. \ref{Eq:Lopsidedness} is that, in the absence of noise, $0\le A_{L}\le 1$.  And, while it is not a factor in this particular study, $A_{L}$ is an integrated quantity making it relatively robust against noise in the profile.

In the lopsidedness equation, one can replace $v_{\rm{sys}}$ in the integral with any velocity in the profile.  As noted in \citet{Deg2020}, there exists a `folding' velocity such that $A_{L}(v_{\rm{equal}})=0$.  They used this idea to introduce a `velocity offset' asymmetry given by
\begin{equation}
    A_{\rm{vo}}=\frac{2|v_{\rm{equal}}-v_{\rm{sys}}|}{w}~,
\end{equation}
where $w$ is the width of the profile.  The factor of 2 sets the limits on this quantity as $0 \le \Delta v \le 1$.  \citet{Manuwal2021} used this same statistic in their study of the \textsc{EAGLE} simulation and noted that it is similar to, but slightly different from the flux-weighted mean velocity used in studies like \citet{Reynolds2020}.  Due to the similarity between $v_{\rm{equal}}$ and the flux weighted mean velocity, $v_{\rm{fw}}$, we have opted to set
\begin{equation}
    A_{\rm{vo}}=\frac{2|v_{\rm{fw}}-v_{\rm{sys}}|}{w}~,
\end{equation}
for the rest of this study.

A third method of quantifying the asymmetry of a velocity profile is the channel-by-channel asymmetry $\mathcal{A}$.  This statistic, which was introduced simultaneously by both \citet{Reynolds2020} and \citet{Deg2020}, calculates the asymmetry by looking at pairs of channels on either side of some folding velocity.  It is given by
\begin{equation}\label{Eq:Asymmetry}
    \mathcal{A}_{v_{\rm{fold}}}=\frac{\sum_{i=1}^{N}|F(v_{l,i})-F(v_{h,i})|\delta v}{\sum_{i=1}^{N}(F(v_{l,i})+F(v_{h,i}))\delta v}~,
\end{equation}
where $v_{\rm{fold}}$ is the folding velocity, $v_{l,i}=v_{\rm{fold}}-i\delta v$ and $v_{h,i}=v_{\rm{fold}}+i\delta v$, $\delta v$ is the channel width, $F(v)$ is the flux at a specific channel, and $N$ is the total number of channel pairs within the limits of the profile. To be very clear, Eq. \ref{Eq:Asymmetry} is applicable at any velocity.
Like $A_{L}$, $\mathcal{A}$ will vary as a function of velocity, but unlike the lopsidedness, the channel-by-channel minimum is not always zero within a profile.  This raises the question of whether $\mathcal{A}$ should always be calculated at $v_{\rm{fold}}=v_{\rm{sys}}$ or at the velocity that minimizes the asymmetry, $v_{\rm{min}}$. In this work, we use the minimal asymmetry, $\mathcal{A}(v_{\rm{min}})=\mathcal{A}_{\rm{min}}$, as it avoids uncertainties in the calculation of $v_{\rm{sys}}$ propagating into uncertainties in the channel-by-channel asymmetry.



\section{General Trends}\label{Sec:GenTrends}

The advantage of working with cosmological simulations in general, and with the \simba\ simulation in particular, is the ability to compare intrinsic properties of a galaxy to the measured asymmetry. Given the range of physical processes that can give rise to morphological and dynamical asymmetries in galaxies, some of the properties that could be correlated with asymmetry are the mass (\hi\ and stellar), 3D distance to the nearest neighbour ($D_{nn}$) where any galaxy, central or satellite, within the caesar catalogue is considered as a potential neighbour, merger number up to $z$~=~1, {the number of dynamical times since the most recent merger ($T_{\mathrm{dyn}}$), \hi\ gas fraction ($f_{\rm{HI}}=M_{\rm{HI}}/M_{*}$), the specific star formation rate (sSFR), and the degree of rotation support, $\kappa$.  This is quantified as the fraction of kinetic energy ($K$) invested in ordered rotation ($\kappa$) as rotation roughly traces morphology (low rotation [$<0.5$] for irregular/elliptical galaxies and high rotation [$>0.7$] for disky galaxies, as defined in \citealt{Sales2012}):
\begin{equation}
    \kappa = \frac{K_{\rm rot}}{K} = \frac{1}{K} \sum \frac{1}{2} m \left(\frac{j_{\rm z}}{r}\right)^{2},
\end{equation}
where $j_{\rm z}$ is the specific angular momentum perpendicular to the disc, ${K_{\rm rot}}$ is kinetic energy in ordered rotation, and $m$ is mass enclosed in radius $r$.

The calculation of the number of dynamical times since the most recent merger is done using a number of approximations.  The redshift since the most recent merger is recorded for each galaxy in \simba, which can be converted to time, $t_{\mathrm{recent}}$.  Then the number of dynamical times since the most recent merger is simply:
\begin{equation}
    T_{\mathrm{dyn}}=\frac{t_{\mathrm{recent}}}{P}~,
\end{equation}
where $P$ is the period of the galaxy.  The period can be approximated by calculating $R_{\hi}$ using the \hi\ size-mass relation of \citet{Wang2016}, and calculating $V_{\hi}$ using the velocity-mass relation derived in \citet{Lewis2019}. Then the approximate period is simply $P=2\pi R_{\hi}/V_{\hi}$.

In order to investigate dependencies on these properties, we compare the average asymmetry statistics, as well as their dispersions, to many of these properties in Fig. \ref{Fig:GeneralTrends}. We have chosen to plot the dispersions rather than the uncertainty in the mean asymmetries in order to highlight the large range of asymmetry measurements in each bin. This is seen by the  large asymmetry dispersion for every measurement. The bins themselves were chosen for each property to ensure adequate statistics per bin and to probe the appropriate parameter space per property. The large asymmetry dispersions are unsurprising given that many different processes may generate asymmetric profiles, and, as shown in \citet{Deg2020}, strongly asymmetric galaxy morphologies can still have symmetric profiles. This was also seen by \citet{Bilimogga2022} who found that the \hi\ morphological (2D) asymmetries and profile (1D) asymmetries of galaxies in the \eagle\ simulation \citep[][]{Schaye2015,Crain2015} were uncorrelated. This also means that the correlations between galaxy properties and asymmetry statistics tend to be somewhat weak.

In Fig. \ref{Fig:GeneralTrends} the channel-by-channel asymmetry (magenta lines), $\mathcal{A}_{\rm{min}}$, is systematically larger than all other statistics.  This is unsurprising as the profiles show significant confusion (as noted in Sec. \ref{Sec:Profiles}).  $A_{\rm{vo}}$ (blue line) is systematically lower than the other measures.  Again, this is unsurprising as it takes a significant amount of flux to generate large offsets between the systemic and flux-weighted mean velocities.  The lopsidedness (red), $A_{L}$, \textbf{lies in between the two other measurements}.

There are some possible trends in Fig. \ref{Fig:GeneralTrends}, but the large asymmetry dispersions make such trends difficult to identify.  One method of highlighting trends in asymmetry is to look at the fraction of galaxies with asymmetry levels above some limit.  That is, plotting $f(A>\mathrm{Lim}_A)$, where $A$ is a particular statistic and $\mathrm{Lim}_{A}$ is the limiting value for that particular statistic.  This is a fairly widespread practice when exploring lopsidedness \citep{Espada2011,Bok2019}.  For that statistic a limit of $\mathrm{Lim}_{A_{L}}=0.12$ is equivalent to $A=1.26$ in the formulation used by \cite{Espada2011} and \cite{Bok2019}.  This value corresponds to the 2$\sigma$ deviation of the asymmetry distribution of isolated galaxies in the sample of \cite{Espada2011} and has been generally used as a dividing line between symmetric and lopsided profiles.

\begin{figure*}
\centering
    \includegraphics[width=150mm]{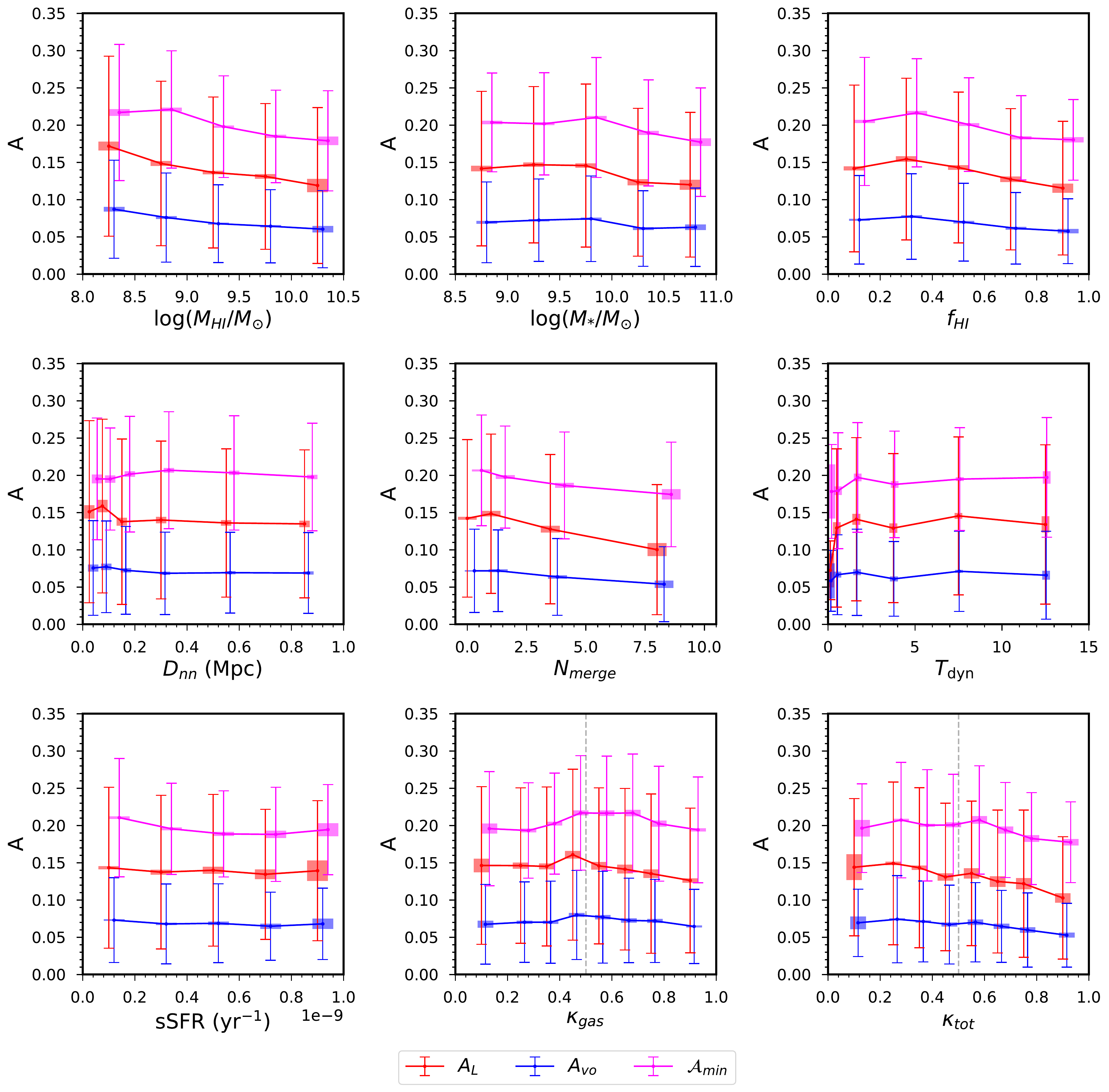} 
\caption{The relationship between different asymmetry measurements and various \simba\ galaxy properties.  The red, blue, and magenta lines are for the lopsidedness, $A_L$, velocity offset, $A_{\rm{vo}}$, and the  channel-by-channel asymmetry, $\mathcal{A}_{\rm{min}}$. The blue and magenta points have been given a slight offset from the red points to improve readability.  The vertical error bars are the dispersion of each particular asymmetry statistic in the bins, while the colored boxes show the uncertainty on the mean}.  The dashed vertical line in the $\kappa_{\rm{gas}}$ and $\kappa_{\rm{tot}}$ panels (bottom row) separates the non-rotators from the rotators.
  \label{Fig:GeneralTrends}
\end{figure*}

Figure \ref{Fig:GeneralTrends} shows that the average value of each of the asymmetry statistics is different.  Therefore, it is not appropriate to use the same limit for each statistic.  Ideally we would derive limits for each statistic using a similar sample to \citet{Espada2011}, but that is beyond the scope of this paper.  Alternatively, it is possible to use the $A_{L} = 0.12$ limit and  to derive limits for $A_{\rm{vo}}$ and $\mathcal{A}_{\rm{min}}$ for this sample in two distinct ways.  The first way is to do a linear fit to the $A_{L}-A_{\rm{vo}}$ and $A_{L}-\mathcal{A}_{\rm{min}}$ relations and find the value corresponding to $A_{L}=0.12$.  The second method is to find the value of $A_{\rm{vo}}$ and $\mathcal{A}_{\rm{min}}$ that keeps the ratio $R=N(A\ge \mathrm{Lim}_{A})/N(A<\mathrm{Lim}_{A})$ constant. Figure \ref{Fig:Lop-AysmCorr} shows the correlations between the different statistics for the full sample of galaxies.  It is clear that the limits of the best-fit line method and the ratio method are approximately the same.  For simplicity all analysis using the fraction above some limit will use limits determined by the ratio method.  That is, $\mathrm{Lim}_{A_{L}}=0.12$, $\mathrm{Lim}_{A_{\rm{vo}}}=0.06$, and $\mathrm{Lim}_{\mathcal{A}_{\rm{min}}}=0.19$ are used in this work. 


\begin{figure*}
\centering
    \includegraphics[width=110mm]{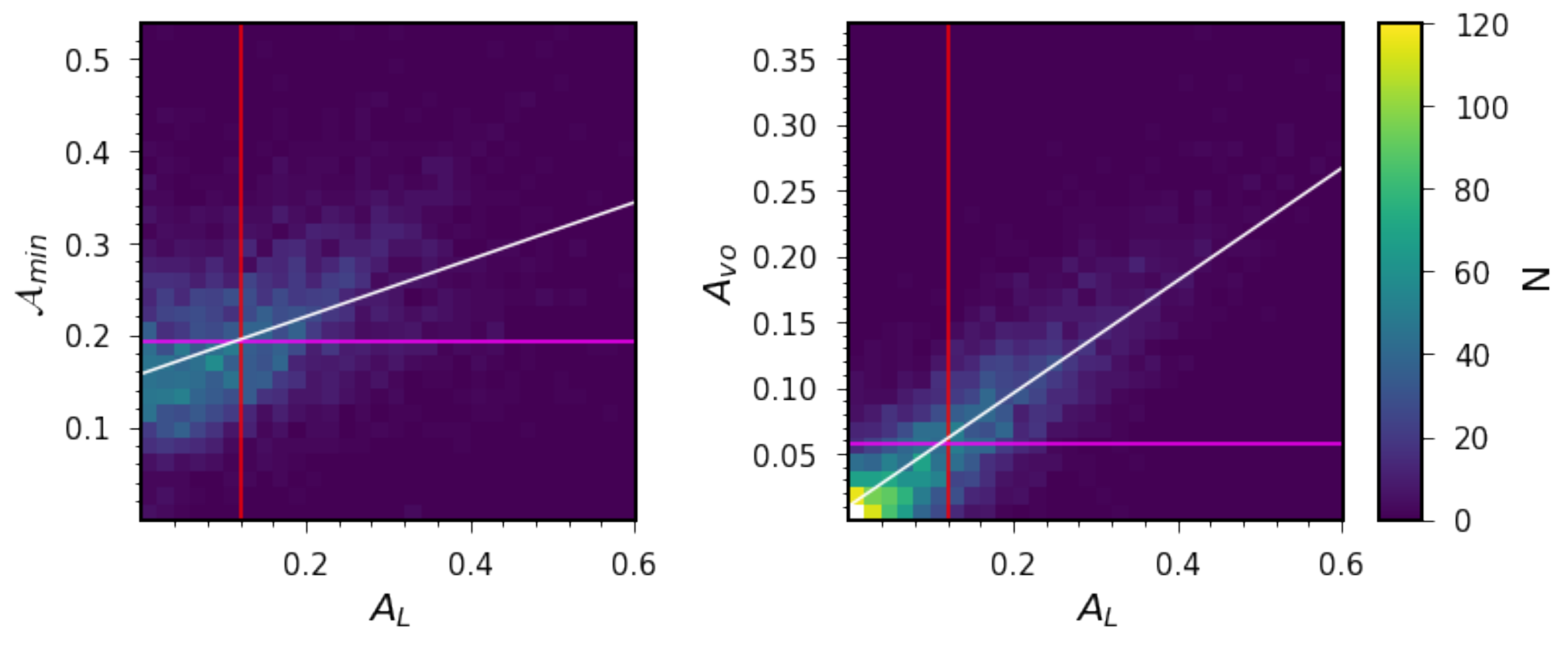} 
\caption{The correlations between the lopsidedness and channel-by-channel asymmetries (left panel) and lopsidedness and $A_{vo}$ asymmetries (right).  The colours show the number density of galaxies in the particular parameter space with blue being low and yellow being high.  The white line shows the best-fit relation between the statistics, the vertical red line shows the lopsidedness limit $\mathrm{Lim}_{A_{L}}=0.12$, and the horizontal magenta lines show the limits required to keep the ratio of asymmetric/symmetric galaxies constant for the other two statistics. In the right hand panel, the best fit line appears slightly low.  This is due to both the larger numbers of galaxies at low $A_{L}$ and $\mathcal{A}_{\rm{min}}$ as well as a diffuse population of galaxies with large $A_{L}$ and low $\mathcal{A}_{\rm{min}}$.}
  \label{Fig:Lop-AysmCorr}
\end{figure*}

Armed with appropriate limits, it is possible to now examine the fraction of `asymmetric' galaxies as determined by these limits.  The general trends for each are shown in Fig. \ref{Fig:General_FracBelow}.  The uncertainties for the fractional analysis are the standard Poisson errors based on the total number of galaxies in each bin.  In order to quantitatively determine whether any of the trends are significant, the Spearman rank coefficient and corresponding $p$-value are calculated for each statistic in each variable.  The Spearman rank coefficient measures the strength of any monotonic trends \citep{Spearman1904}.  It has limits of $-1\le\rho\le 1$, where $1$ indicates an increase, $0$ shows no relation, and $-1$ is a general decrease. The $p$-values measure the significance of the Spearman rank coefficient.  We have adopted the standard definitions where $p<0.01$  is significant, $0.01\le p < 0.05$ is marginally significant, and $p\ge 0.05$ is not significant.  For simplicity we use the \textsc{SciPy} \citep{SciPy} implementation of the Spearman rank coefficient calculation.

\begin{figure*}
\centering
    \includegraphics[width=150mm]{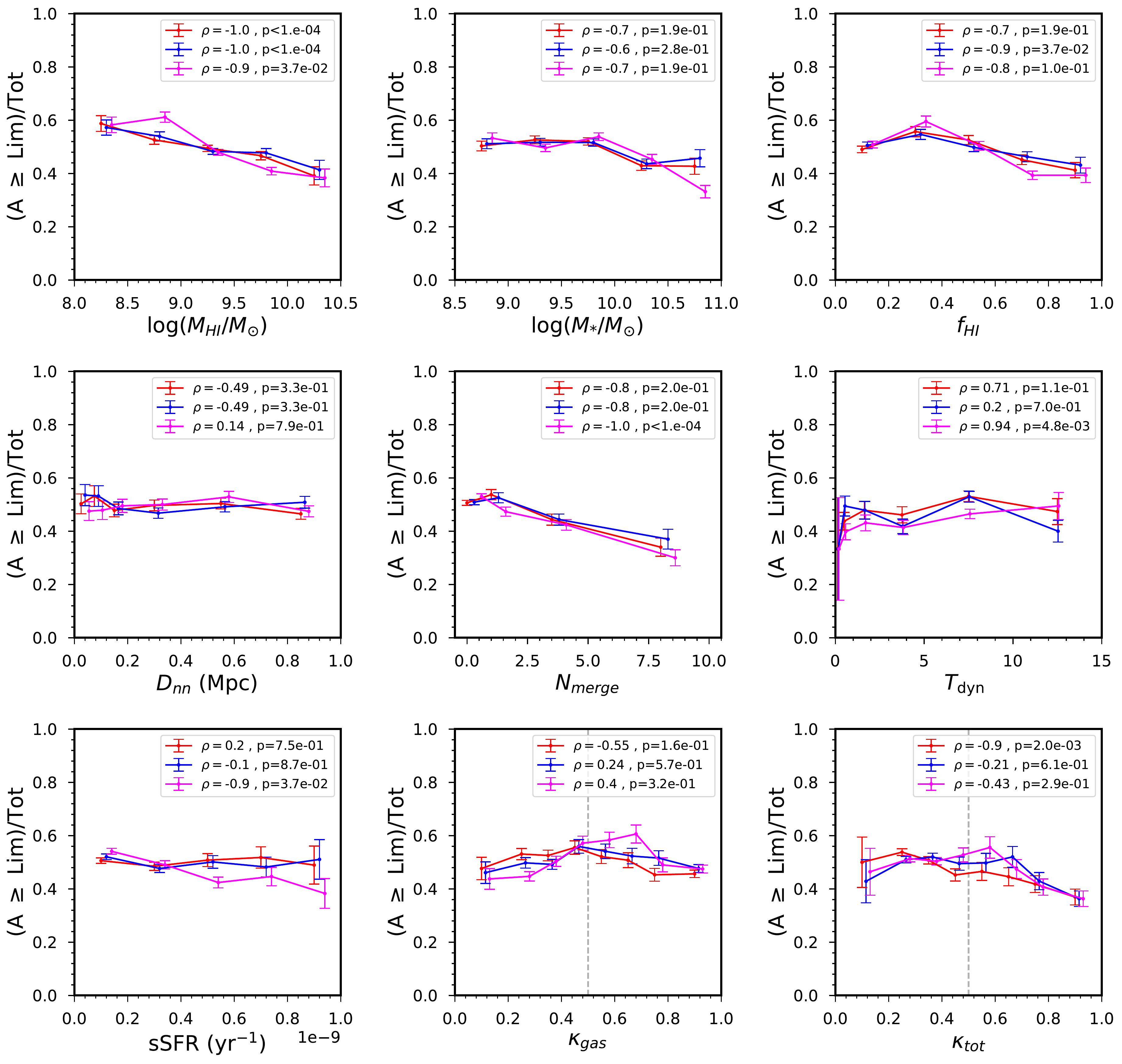} 
\caption{The fraction of galaxies having an asymmetry measurement above the fiducial limits (as shown in Fig. \ref{Fig:Lop-AysmCorr} as $A_{L}\geq 0.12$, $A_{vo}\geq 0.06$, and $\mathcal{A}_{min}\geq 0.19$) in each bin compared to the various \simba\ galaxy properties.  The coloured lines and offsets in all panels and the dashed vertical lines in the bottom row of panels are as in Fig. \ref{Fig:GeneralTrends}.  The vertical error bars are the standard rms errors using the total number of galaxies in each bin.  The Spearman rank coefficients and their associated $p$ values are given in the legends.  }
  \label{Fig:General_FracBelow}
\end{figure*}

Examining the trends and Spearman rank coefficients seen in Fig. \ref{Fig:General_FracBelow}, the strongest relation with the lowest $p$-value is the asymmetry- $M_{\rm HI}$ relation. The fraction of asymmetric galaxies, using all statistics, $A_{L}$, $A_{vo}$, and $\mathcal{A}_{\mathrm{min}}$, decreases with increasing \hi\ mass.

The asymmetry-stellar mass relations show a somewhat similar set of correlations ($\rho<0$), but they are much weaker than the relationship with $M_{\rm HI}$.  The high $p$-values indicate that the $\rho$ values are not significant; i.e. the data is consistent with there not being a global relationship between asymmetry and stellar mass. 
Given the strong $M_{\hi}$ relationship and the lack of a global stellar mass relationship, it is unsurprising that there is perhaps a weak trend in the in the gas fraction panel.  However, the gas fraction $p$ values do not indicate that such a trend is significant. Nonetheless, in Sec. \ref{Sec:Watts}, when the $\hi$ mass is controlled, we see that there is a secondary trend with gas fraction.

There is no clear evidence for a trend with distance to nearest neighbour.  This is different to the results of \citet{Bok2019}, where they found that isolated galaxies tended to have lower lopsidedness values on average compared to close pairs, while our results show nearly constant asymmetries for separations up to 1 Mpc. There is perhaps a small increasing trend in lopsidedness for $D_{nn}<200$~kpc, with a peak at $\sim 100$~kpc for $A_{vo}$ and $A_{L}$ in Fig.~\ref{Fig:GeneralTrends} but the Spearman's rank test does not support a significant trend in Fig.~\ref{Fig:General_FracBelow}. However, given that there are a variety of processes that could give rise to asymmetries and also that galaxy-galaxy interactions are only likely to cause changes to galaxy morphology at relatively short distances (our first $D_{nn}$ bin is $< 100$ kpc), our result is perhaps not surprising. We also compared the average asymmetry values for $D_{nn}<100$ kpc and $D_{nn}>1$ Mpc but there was no significant difference between them. 
There are a few possible factors which could be contributing to washing out a possible trend. If the simulation catalogue does not distinguish two galaxies that are in the process of merging as separate objects, then the resulting $D_{nn}$ will refer to the distance to a third neighbour which could have a wide range of values. However, the merger object will likely have a high asymmetry, thereby inflating the average asymmetry at higher $D_{nn}$. This is illustrated in the middle panel of Fig.~\ref{Fig:SampleProfiles} for gal4718. This galaxy is listed as having a very large distance to nearest neighbour of $D_{nn}=1.7$~Mpc, however the \hi\ image shows what is likely to be two systems merging and the shape of the \hi\ profile indicates the same. 
Another contributing factor to inflating the average asymmetry for larger $D_{nn}$ values could be source confusion due to cubes containing \hi\ emission from multiple neighbouring galaxies; similarly to real observations, nearby sources will be included in the data cubes depending on the size specifications used in cube generation. This situation is illustrated in the last row of Fig.~\ref{Fig:SampleProfiles} where in the catalogue gal2322 is listed with a distance to nearest neighbour of $D_{nn}=0.7$~Mpc but the neighbour is clearly seen in the \hi\ image and as confusion in the \hi\ profile. In this particular case, our methods successfully disentangled the two profiles, but this is not always possible. This will occur in real observational data as well.



In Figs. \ref{Fig:GeneralTrends} and \ref{Fig:General_FracBelow}, there doesn't appear to be a strong correlation between the sSFR and the asymmetry. However, as with the gas fraction, in Sec. \ref{Sec:Watts} where the $\hi$ mass is accounted for, a secondary trend with sSFR appears.

There may also be a weak correlation with rotation.  Both the gas rotation and total rotation suggest that rotating galaxies tend to be more symmetric than non-rotating galaxies. Moreover, there is an apparent peak in the asymmetry for $\kappa_{gas}\approx 0.5$, and all the asymmetry measurements decrease for $\kappa_{tot}\ge 0.5$.  This is particularly interesting as that is roughly the dividing line between rotators and pressure supported systems.  Table \ref{Tab:Rotation} quantifies the trends in the pressure-supported ($\kappa<0.5$) and rotating ($\kappa\ge0.5$) systems using the Spearman rank coefficients for the fraction of symmetric galaxies (as is done in Fig. \ref{Fig:General_FracBelow}) for the entire population.  These quantifications show that the trends are generally strongest for the $\kappa_{tot}\ge0.5$ population.  In other words, intermediate objects ($\kappa \sim 0.5$) have systematically larger asymmetries than rotating objects.  One possible reason for this rotation trend is that intermediate objects may have undergone some sort of interaction to move them from pressure-supported or rotation supported.  However, there may be other drivers of this trend, included total mass, gas fraction, etc.  Fully exploring this trend will be the subject of future work.

\begin{table}
    \centering
    \begin{tabular}{|c|c|c|}
        \hline 
         & $\kappa_{\rm{gas}}<0.5$ & $\kappa_{\rm{gas}}\ge 0.5$ \\
         \hline 
        $A_{L}$ &$\rho=0.8,~p=0.2$ & $\rho=-0.8,~p=0.2$\\ 
        $A_{\rm{vo}}$ & $\rho=0.8,~p=0.2$& $\rho=-1.0,~p<10^{-4}$\\ 
        $\mathcal{A}_{\rm{min}}$ & $\rho=1.0,~p<10^{-4}$& $\rho=-0.8,~p=0.2$ \\ 
        \hline 
        & $\kappa_{\rm{tot}}<0.5$ &$\kappa_{\rm{tot}}\ge 0.5$ \\
        \hline
        $A_{L}$ &$\rho=0.4,~p=0.6$ & $\rho=1.0,~p<10^{-4}$\\
        $A_{\rm{vo}}$ & $\rho=-0.4,~p=0.6$& $\rho=0.8,~p=0.2$\\ 
        $\mathcal{A}_{\rm{min}}$ &$\rho=-0.8,~p=0.2$ & $\rho=1.0,~p=10^{-4}$ \\ 
        \hline 
    \end{tabular}
    \caption{The Spearman rank coefficients for the pressure supported population $\kappa < 0.5$ and rotation supported population $\kappa\ge 0.5$.}
    \label{Tab:Rotation}
\end{table}

There is no clear correlation between the asymmetry and the number of dynamical times since the most recent merger, except possibly for $T_{\mathrm{dyn}} \le 1$.  However, as there are only a few objects that have their most recent merger within this time frame, it is difficult to draw a quantitative conclusion in the low time limit.  The roughly constant asymmetry when mergers have occurred longer ago than two dynamical times suggests that other drivers of asymmetry may be limiting the ability of the galaxy to settle back into a symmetric morphology.

Finally there does appear to be some correlation with the number of mergers.  While the $p$ values are generally not quite as small as those seen for the $M_{\hi}$ relation, this is driven mostly by the down-turn in the asymmetry measurement at $N_{\rm{merge}}=0$. As with the $M_{\rm HI}$ relation, the asymmetry generally decreases as a function of the number of mergers. The merger relation is likely another version of the $M_{\rm HI}$ relation as the galaxies with the most gas mass typically have undergone the largest number of mergers.

In order to explore the trend with mergers in greater detail, the galaxies that have undergone a merger within $z\le 1$ are separated from those that have not in Fig.~\ref{Fig:MergerCompTrends}.  The panels show the relationship between the profile asymmetries and $M_{\rm HI}$, $M_{*}$, $f_{\rm HI}$, and $D_{nn}$ for both unmerged (solid lines) and merged (dashed lines) galaxies.  
Galaxies that remain unmerged show stronger trends with $A_{L}$ and $A_{vo}$ than those that have undergone a merger.  While somewhat counter-intuitive, this result shows the effect of mergers on profile asymmetries.  They wash out the \hi\ mass dependence. 
Separating into merged and unmerged samples results in different trends in average asymmetry vs $D_{nn}$. The hint of an increase in asymmetry around $D_{nn}\sim100$~kpc seen in Fig.~\ref{Fig:GeneralTrends} seems to be driven by the unmerged galaxies as seen in the bottom left panel of Fig.~\ref{Fig:MergerCompTrends}. This is likely due to many of the nearby unmerged galaxies being affected by tidal forces or being in the early process of merging, but not yet fully merged. There is also an offset in the fractions of symmetric galaxies between the merged and unmerged samples for lower $D_{nn}$ values, with the largest difference at around $D_{nn}\sim100$ kpc, as shown in the bottom right panel of Fig.~\ref{Fig:MergerCompTrends}. For larger distances to nearest neighbour the two samples tend towards similar fractions of $\sim50$ percent.

\begin{figure*}
\centering
    \includegraphics[width=110mm]{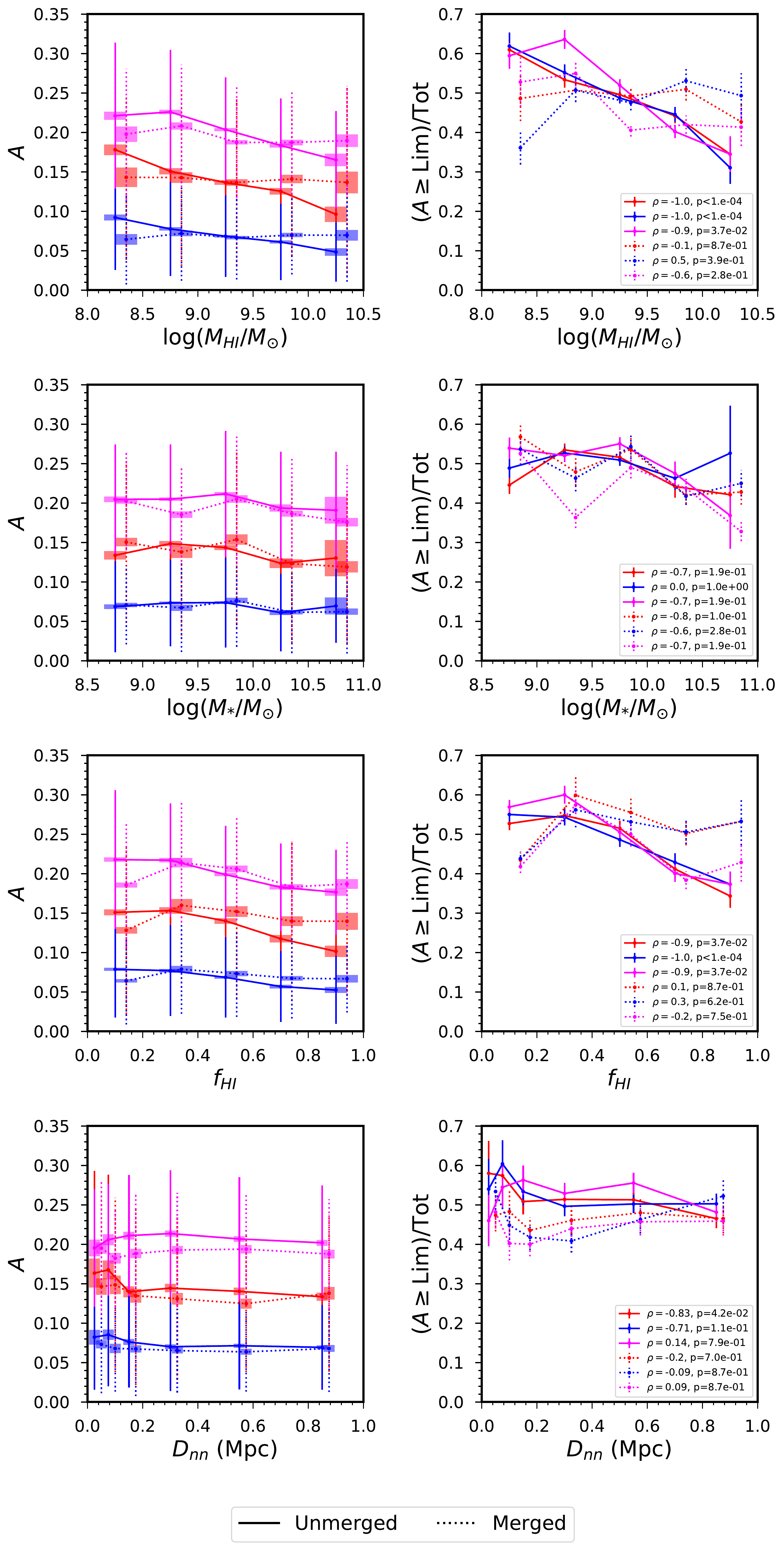} 
\caption{The relationship between different asymmetry measurements and various \simba\ galaxy properties for galaxies that have undergone a merger at $z<1$ (solid lines) and those that have not (dotted lines). The subsamples have been offset along the $x$-axis for clarity. The left-hand column shows the asymmetry measurements, while the right hand column shows the fraction above the fiducial asymmetry limits.  Colours and error bars in the left-hand panels are as in Fig. \ref{Fig:GeneralTrends}, while the error bars in the right-hand panels are as in Fig. \ref{Fig:General_FracBelow}.}
  \label{Fig:MergerCompTrends}
\end{figure*}

Given the \hi\ mass dependence seen in Figs. \ref{Fig:GeneralTrends}-\ref{Fig:General_FracBelow}, it is also reasonable to divide the profiles into a low \hi\ mass ($\log(M_{\hi}/M_{\odot}) < 9.4$) sample and a high \hi\ mass sample ($\log(M_{\hi}/M_{\odot}) > 9.4$). This particular mass limit is chosen based on the \citet{Watts2021} analysis (which is discussed in much greater detail in Sec. \ref{Sec:Watts}).  The resulting asymmetry measures and symmetric/total galaxy ratios are shown in Fig. \ref{Fig:MassCompTrends}. The offset between the solid and dashed lines is a reflection of the overall \hi\ mass trend. For instance, the low $M_{\rm HI}$ galaxies show a larger change in profile asymmetries with stellar mass than high $M_{\rm HI}$ galaxies.  

The high $M_{\rm HI}$ (more gas rich) galaxies maintain relatively constant asymmetry values as a function of stellar mass. However, the low $M_{\rm HI}$ (gas-poor) galaxies have higher asymmetries than their gas-rich counterparts at low stellar masses and cross over to lower asymmetry values around $\log(M_{*}/M_{\odot})\sim10$.
Similarly, the high $M_{\rm HI}$ galaxies do not show a trend with the number of mergers, while the lower $M_{\rm HI}$ galaxies do show a trend according to their Spearman rank values.  This suggests that repeated mergers tend to `smooth' the gas distribution for lower mass galaxies, making it more symmetric.  But, for higher $M_{\rm HI}$ galaxies, the mergers have less of an effect on the overall profile asymmetry.  
For distance to nearest neighbour, the asymmetry values remain relatively constant for all $D_{nn}$ for both low and high $M_{\rm HI}$ galaxies, except for at the shortest distances ($\sim100$ kpc) where there might be a hint of different trends for the two samples. But, given the large dispersions, issues of confusion, and how that relates to the $D_{nn}$ measurement, it is difficult to make any firm conclusions about these trends.  

Finally, there does not appear to be a difference in the asymmetry trends for low and high mass galaxies and the sSFR.

\begin{figure*}
\centering
    \includegraphics[width=110mm]{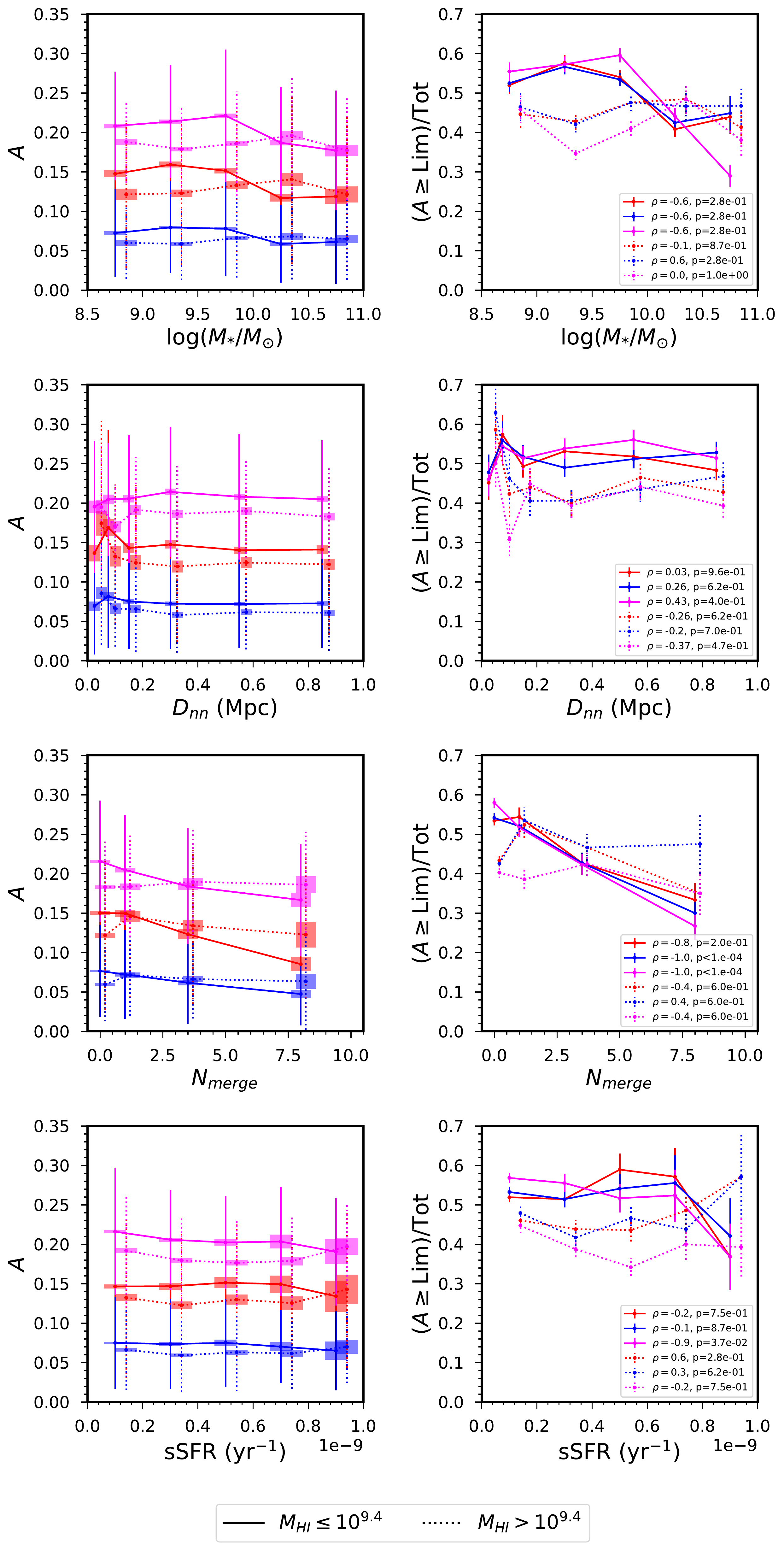} 
\caption{The relationship between different asymmetry measurements and various \simba\ galaxy properties for $\log(M_{\hi}) < 9.4$ (solid lines) and  $\log(M_{\hi}) > 9.4$ (dotted lines). The subsamples have been offset along the $x$-axis for better clarity. The left-hand column shows the asymmetry measurements, while the right hand column shows the fraction above the fiducial asymmetry limits.  Colours and error bars in the left-hand panels are as in Fig. \ref{Fig:GeneralTrends}, while the error bars in the right-hand panels are as in Fig. \ref{Fig:General_FracBelow}.}
  \label{Fig:MassCompTrends}
\end{figure*}

\section{Impact of \hi\ mass fraction and sSFR on symmetry}\label{Sec:Watts}

\begin{figure*}
\centering
    \includegraphics[width=0.99\linewidth]{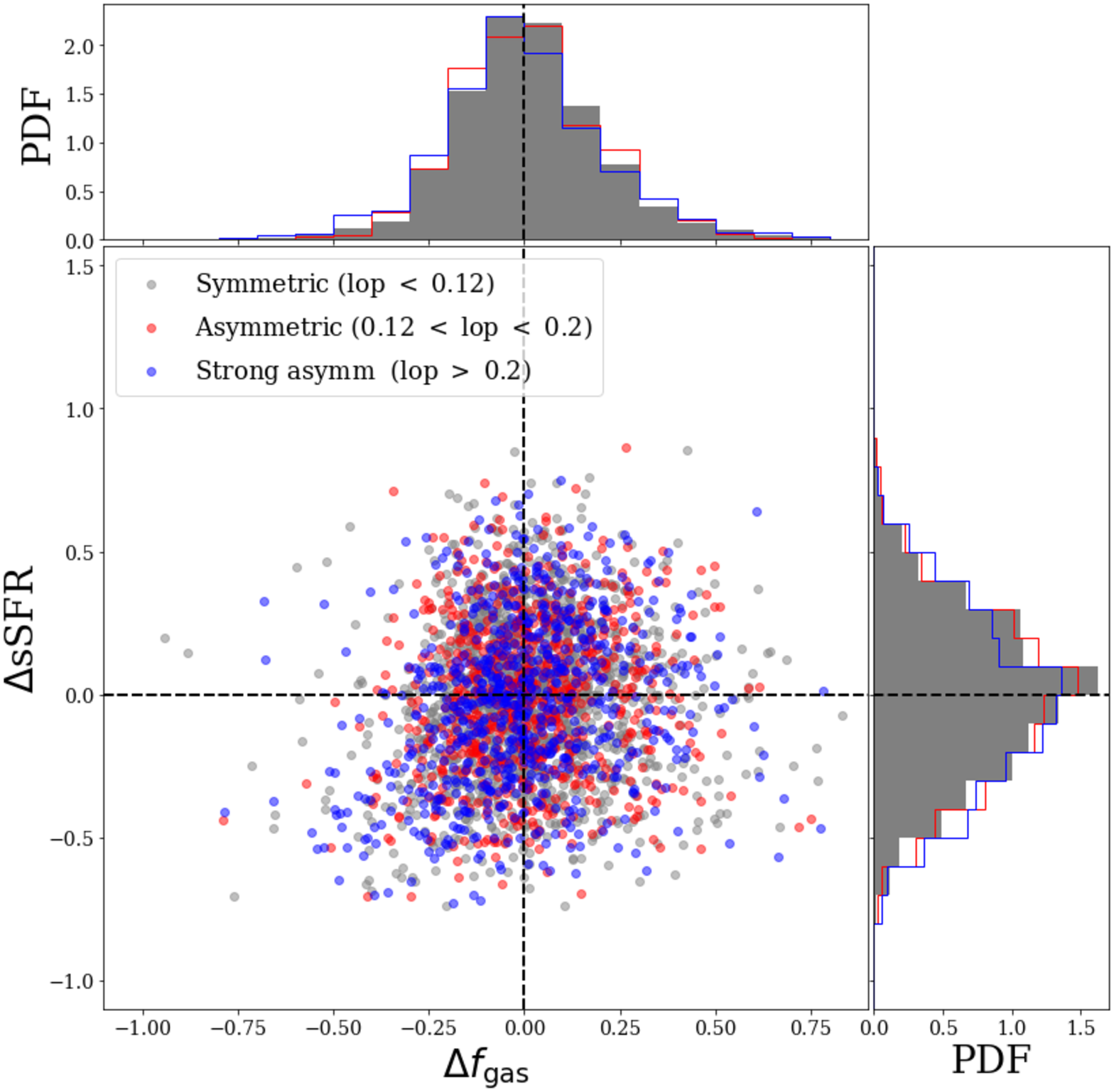} 
\caption{The $\Delta f_{\rm gas}$--$\Delta$sSFR parameter space for \simba\ galaxies. The central panel shows the location of galaxies in the parameter space, with symmetric (lopsidedness~$<$~0.12) galaxies in grey, strongly asymmetric (lopsidedness~$>$~0.2) in blue, and intermediate asymmetric galaxies in red. The corresponding histograms in each parameter space are given in the same colour scheme. The plot layout is akin to fig.~4 of \citet{Watts2021}.}
  \label{Fig:SimbaWattsSymAsym}
\end{figure*}

\begin{figure*}
\centering
    \includegraphics[width=0.625\linewidth]{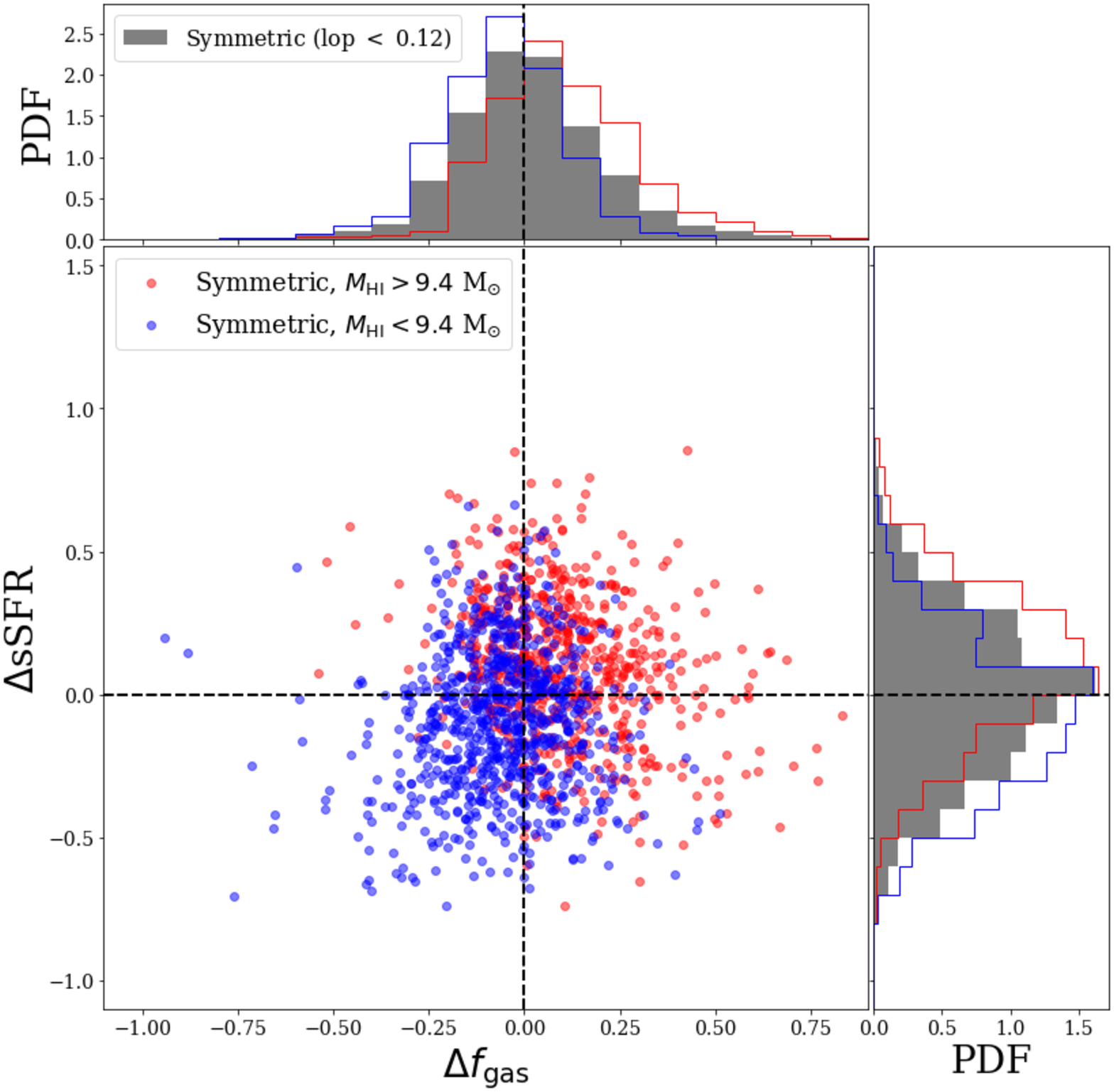} 
    \includegraphics[width=0.625\linewidth]{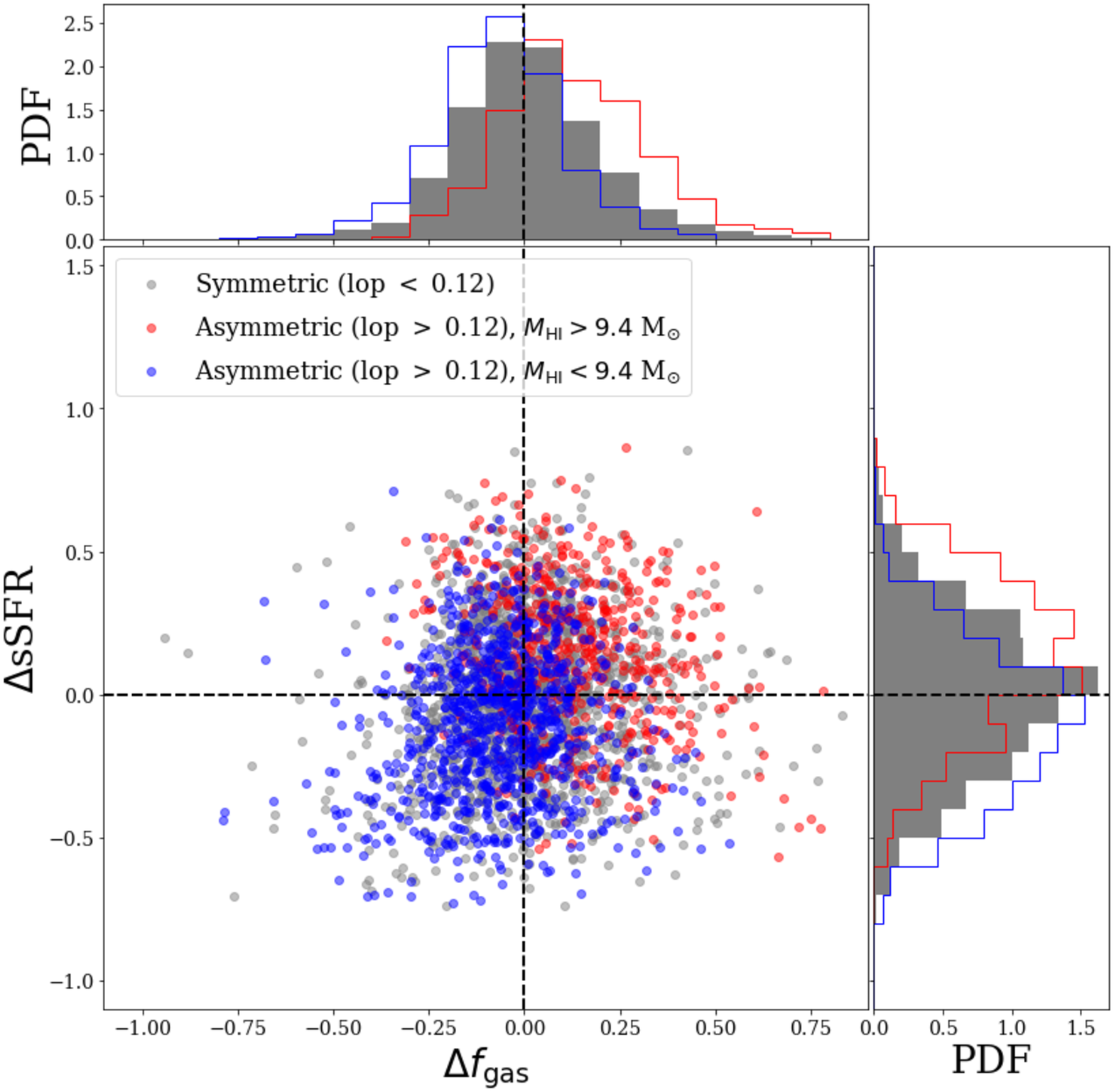} 
\caption{The $\Delta f_{\rm gas}$--$\Delta$sSFR parameter space for \simba\ galaxies, as in Fig.~\ref{Fig:SimbaWattsSymAsym}, but now comparing the high and low $M_{\rm HI}$ galaxies (red and blue respectively) for symmetric (top) and asymmetric (bottom) galaxies. The grey points in the bottom panel are the symmetric galaxies from the top panel.}
  \label{Fig:SimbaWattsMassCuts}
\end{figure*}

\begin{figure*}
\centering
    \includegraphics[width=0.99\linewidth]{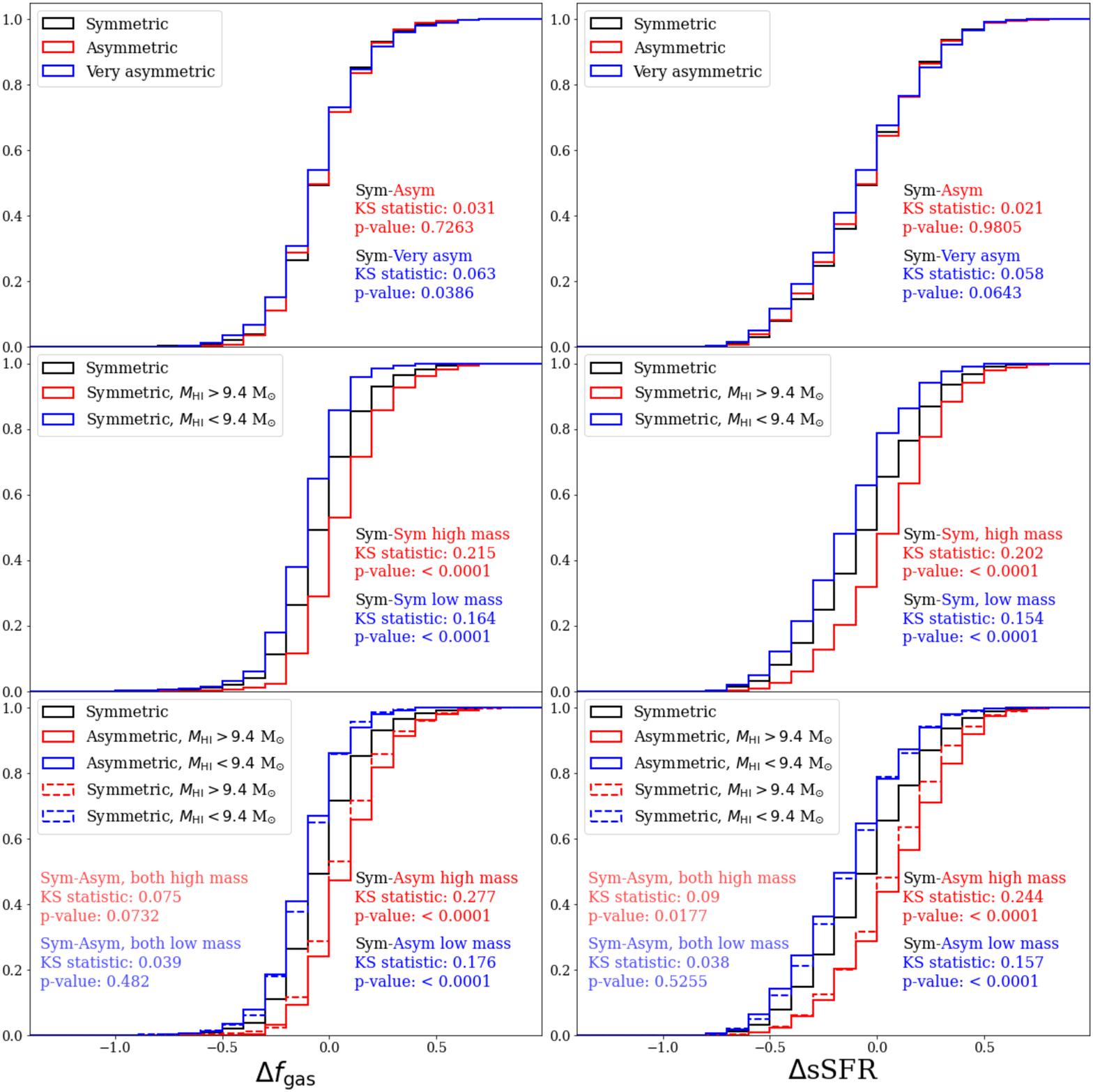} 
\caption{Cumulative histograms for the samples presented in Figs.~\ref{Fig:SimbaWattsSymAsym} and \ref{Fig:SimbaWattsMassCuts}. Left sided panels give the distributions for the $\Delta f_{\rm gas}$ parameter, and right for $\Delta$sSFR. The 2-sample KS statistics and corresponding p-values between the symmetric population (black) and asymmetric (red), strongly asymmetric (blue) distributions are given on their respective panels. In the bottom row panels, we also compare the asymmetric (solid) and symmetric (dashed) high \hi\ mass (red) and low \hi\ mass (blue) subsamples. We see stronger evidence to reject the null hypothesis that the distributions are the same for $\Delta f_{\rm gas}$, and when $M_{\rm HI}$ limits are considered. The low and high mass populations of symmetric and asymmetric galaxies cannot be rejected so readily, especially for the $\Delta$sSFR parameter.}
  \label{Fig:Cumulative}
\end{figure*}

\begin{table*}
    \centering
    \begin{tabular}{|p{0.4\linewidth} | p{0.54\linewidth}|}
        \hline 
        \textbf{Populations} & \textbf{Trends} \\
        \hline 
        Symmetric, intermediate asymmetry and strongly asymmetric \simba\ galaxies (Fig.~\ref{Fig:SimbaWattsSymAsym}, top row Fig.~\ref{Fig:Cumulative}). &
        Weak trend seen for strongly asymmetric galaxies to have lower HI mass fractions. Insignificant difference in $\Delta$sSFR. In agreement with \citet{Watts2021}. \\
        \hline 
        Symmetric galaxies with high $M_{\rm HI}$, and symmetric galaxies with low $M_{\rm HI}$ (top panel Fig.~\ref{Fig:SimbaWattsMassCuts}, middle row Fig.~\ref{Fig:Cumulative}). &
        High $M_{\rm HI}$ symmetric galaxies have high HI mass fractions and sSFR, and vice versa.\\
        \hline 
        All asymmetric galaxies with high $M_{\rm HI}$, and all asymmetric galaxies with low $M_{\rm HI}$ (bottom panel Fig~\ref{Fig:SimbaWattsMassCuts}, bottom row Fig.~\ref{Fig:Cumulative}). &
        Similarly to above, high $M_{\rm HI}$ asymmetric galaxies have high HI mass fractions and sSFR, and vice versa. \\
        \hline 
        Symmetric galaxies with high $M_{\rm HI}$, and all asymmetric galaxies with high $M_{\rm HI}$ (bottom row Fig.~\ref{Fig:Cumulative}, red solid and dashed lines). &
        Weak trend seen in asymmetric high $M_{\rm HI}$ galaxies having higher sSFR than symmetric high $M_{\rm HI}$ galaxies. No significant trend seen in HI mass fraction. \\
        \hline 
        Symmetric galaxies with low $M_{\rm HI}$, and all asymmetric galaxies with low $M_{\rm HI}$ (bottom row Fig.~\ref{Fig:Cumulative}, blue solid and dashed lines). &
        No difference observed in $f_{\rm gas}$ or sSFR for asymmetric/symmetric galaxies with low $M_{\rm HI}$.\\ 
        \hline 
    \end{tabular}
    \caption{Summary of the populations examined and trends found in Section~\ref{Sec:Watts}.}
    \label{Tab:WattsCompSumm}
\end{table*}

We now focus on two specific galaxy properties, the \hi\ gas fraction and specific star formation rate, and how they relate to asymmetry, and draw inspiration from the findings of \cite{Watts2021}. They demonstrated the importance of including gas-poor star-forming galaxies in their analysis of \hi\ profile symmetry, where asymmetric galaxies were typically more gas-poor than symmetric galaxies at fixed stellar mass, with no change in specific star formation rates (sSFR). Like our findings in Section~\ref{Sec:GenTrends}, they also demonstrated that merger activity does not always lead to an asymmetric global \hi\ spectrum. As we can easily obtain gas fractions and sSFR for \simba\ galaxies, we investigate how well our sample compares to the \cite{Watts2021} study of ALFALFA \citep{Haynes2018} and xGASS \citep{Catinella2018} samples. This allows for an exploration of whether asymmetry can trigger or enhance star formation.

\subsection{2D distributions and statistics}

We begin by implementing a similar sample cut as \cite{Watts2021} to more directly compare to their findings. The same fits found by \cite{Janowiecki2020} for the xGASS star forming main sequence (SFMS) were used to select galaxies more star-forming than 1.5$\sigma_{\rm MS}$ below the SFMS, to remove galaxies that have undergone significant suppression of their SFR. That is, we select galaxies above the dashed red line in the top panel of Fig.~\ref{Fig:SimbaSample}, with stellar masses log($M_{*}$/$\mathrm{M}_\odot$)~$>$~9. From this sub-sample, we further separate galaxies into three levels of symmetry based on the lopsidedness measure $A_{L}$ which was also done by \cite{Watts2021}: symmetric ($A_{L}$~$<$~0.12; 1448 galaxies), asymmetric (0.12~$<$~$A_{L}$~$<$~0.2; 732 galaxies), and very asymmetric ($A_{L}$~$>$~0.2; 752 galaxies). We do not remove galaxies outside the stellar mass range of the xGASS sample studied by \cite{Watts2021}, although we note there are no significant differences in our results if we implement such a cut. We note that as shown in Fig.~\ref{Fig:SimbaSample}, the \simba\ SFMS is higher than the xGASS SFMS. The following results do not change significantly for a sample selected above the \simba\ SFMS-1.5$\sigma$ cut (green solid line of Fig.~\ref{Fig:SimbaSample}); such a sample includes more higher stellar mass galaxies, and less lower-mass galaxies, relative to the xGASS-like sample used for the following results.

Next we perform a matched-galaxy offset analysis \cite[see also][]{Ellison2018,Watts2020b}. We bin all galaxies by stellar mass, and match the asymmetric galaxies to symmetric galaxies within the same bin. Care was taken to ensure at least five symmetric galaxies occupy each bin. We note that unlike \cite{Watts2021} we do not also match on signal-to-noise (S/N), as we do not include noise in our \simba\ \hi\ cubes. \cite{Watts2021} stated that matching in S/N had no effect when analysing their ALFALFA sample (1784 galaxies) which had a larger number of sources than their xGASS sample (322 galaxies); our sample size is greater than these two samples combined, evident when comparing Fig.~\ref{Fig:SimbaSample} with fig. 2 of \cite{Watts2021}. 

To study the \hi\ mass fraction ($M_{\rm HI}$/$M_{*}$) and sSFR properties of our sample, we follow the definitions of \cite{Watts2021} of using offsets:

\begin{equation}
    \Delta f_{\rm gas} = {\rm log(}M_{\rm HI}/M_{*})_{\rm asym,sym} - {\rm med[log(}M_{\rm HI}/M_{*})_{\rm sym,match}]
\end{equation}
\begin{equation}
    \Delta {\rm sSFR} = {\rm log(sSFR)}_{\rm asym,sym} - {\rm med[log(sSFR)}_{\rm sym,match}].
\end{equation}

That is, we subtract the median of our symmetric galaxy sample, at a fixed $M_{*}$ from both the symmetric and asymmetric populations. We highlight that our sample includes gas-poor galaxies, unlike the \cite{Watts2021} ALFALFA sample, and so we expect results to resemble that of fig.~6 of \cite{Watts2021} which focused on their xGASS sample, albeit with a significantly larger sample size.

In Fig.~\ref{Fig:SimbaWattsSymAsym} we show the $\Delta f_{\rm gas}$--$\Delta$sSFR parameter space for the \simba\ sample for our three different sub-samples divided in asymmetry value. The dashed lines are set at 0 offsets in both parameters. Above and right of the central panel are density-normalised distributions for each parameter. We next present Fig.~\ref{Fig:SimbaWattsMassCuts}, giving the high and low $M_{\rm HI}$ galaxies found to be symmetric (top panel; 627 and 821 galaxies respectively) and \emph{all} asymmetric galaxies, strongly or otherwise (bottom panel; 524 and 960). We use the same dividing mass limit as before to match \cite{Watts2021}, which provides reasonably sized samples with \simba. 
   
Accompanying these figures is Fig.~\ref{Fig:Cumulative}, where we give the cumulative distributions for $\Delta f_{\rm gas}$ (left-hand side panels) and $\Delta$sSFR (right side). We took two-sample KS (Kolmogorov-Smirnov) tests to determine whether two of these probability distributions differ, comparing either the symmetric sample to the asymmetric, very asymmetric, and high/low-mass symmetric/asymmetric subsamples; or the high/low-mass symmetric and high/low-mass asymmetric subsamples with each other (left side of bottom row panels). The KS statistic and p-value for each of these are displayed on the corresponding panels. In Table~\ref{Tab:WattsCompSumm} we list the populations described above, and summarise the trends for these populations discussed in the following section.  

\subsection{Mass matters: reconciling differences in our subsamples}

First we consider $\Delta f_{\rm gas}$. The asymmetric and strongly asymmetric galaxies tend to have lower \hi\ mass fractions, with this trend more evident for the latter population. While we cannot reject the hypothesis that symmetric and intermediate asymmetry galaxies are drawn from the same population, we can weakly reject it for the symmetric and strongly asymmetric galaxies at a p-value of 0.0386. This agrees with what is shown in fig.~6 of \cite{Watts2021}, who found $p$~=~0.03. Therefore, the \simba\ suite of simulations support the finding that the asymmetric population have on average lower gas fractions (also evident in Fig.~\ref{Fig:MergerCompTrends} for unmerged galaxies), and highlight the importance of \hi-sensitive studies. 

\cite{Watts2021} had found this trend to lower \hi\ mass fractions was further enhanced by the low \hi\ mass asymmetric population, and so we expect the converse to also hold for high \hi\ mass asymmetric galaxies. The bottom-left panel of Fig.~\ref{Fig:Cumulative}, as well as the bottom panel of Fig.~\ref{Fig:SimbaWattsMassCuts}, demonstrates a strong difference in the two populations. Again, this matches our findings in general trends, and so is not surprising that \hi\ mass has an impact here. But before we get ahead of ourselves, we should realise that this may also affect the symmetric population - are they distinct from asymmetric galaxies at the low and high \hi\ mass end? \cite{Watts2021} did not consider it for the symmetric xGASS galaxies due to low sample size, so we explore it here. The top panel of Fig.~\ref{Fig:SimbaWattsMassCuts} and middle-left panel of Fig.~\ref{Fig:Cumulative} shows that the same effect holds for the symmetric subsamples. 

Hence the follow-up question: are the low \hi\ mass symmetric and low \hi\ mass asymmetric galaxies drawn from the same population, and likewise for the high \hi\ mass subsamples? Our cumulative distributions and corresponding KS tests (bottom-left panel of Fig.~\ref{Fig:Cumulative}) suggest not for $\Delta f_{\rm gas}$ at the low \hi\ mass end. The p-value of our KS two-population test is 0.0732 for the high mass subsamples, and 0.4820 for low \hi\ mass galaxies. Therefore, we cannot convincingly reject the null hypothesis that high \hi\ mass symmetric and asymmetric galaxies are the same. It is evident that \hi\ mass and \hi\ gas fraction are major factors for both populations, especially at the low \hi\ mass end. 

When comparing distributions and corresponding KS test results between left and right hand side panels, we often see lower KS statistics and higher p-values for $\Delta$sSFR. The symmetric and intermediate asymmetric samples are indistinct, and only when we consider strongly asymmetric galaxies do we see a weak trend towards lower sSFRs emerge (p-value of 0.0643). There is still a clear difference between the high and low \hi\ mass subsamples, which we note was not evident from the \cite{Watts2021} sample. There is a clear (albeit weak) difference for the high \hi\ mass symmetric and asymmetric populations ($p$~=~0.0177), but again indistinguishable at the low-mass end, where the dominating factor is the \hi\ mass, not whether the galaxy's global profile is symmetric or not.

As summarised in Table~\ref{Tab:WattsCompSumm}, it is clear that the \hi\ mass and \hi\ gas fraction are major factors and drivers of galaxy growth and evolution. For this sample of \simba\ galaxies, only a weak difference can be seen in the sSFR between high \hi\ mass symmetric and asymmetric galaxies. Across the whole sample (no \hi\ mass cuts), only the most asymmetric galaxies deviate weakly in \hi\ gas fraction. The importance of \hi\ mass and \hi\ gas fractions has already been seen to extend beyond asymmetry studies; for example, \cite{Hardwick2021} demonstrated for the xGASS sample that H{\sc i} gas fraction remains the strongest correlated parameter with the scatter of the stellar mass vs. specific angular momentum (Fall) relation. \cite{Pina2021} also found this dependency within a separate nearby disc galaxy sample. These observational findings have also been supported within \simba\ galaxies, where \hi\ accretion history has been hypothesised to be a dominant driver of scatter in the Fall relation (Elson, Glowacki \& Davé, in prep.). 

Overall, we conclude that $\Delta f_{\rm gas}$ showcases distinct differences between each of these subsamples, whereas $\Delta$sSFR is only noticeably diminished for the most asymmetric galaxies, of those which have not been completely quenched. It is important to consider the impact of mass for the whole sample, since for both parameter spaces compared here with the findings of \cite{Watts2021}, the \hi\ mass dominates the trends, particularly for sSFR. Larger, and more sensitive, galaxy samples with upcoming SKA pathfinder HI emission surveys will help add statistics and enable a direct comparison to our \simba\ sample, and extend to higher redshifts.

\section{Conclusions}\label{Sec:Conclusions}

We have laid the groundwork for the beginning of ASymba, a study of galaxy asymmetries for the \simba\ hydrodynamical suite. We constructed a sample of simulated \hi\ cubes through \martini\ and their corresponding 1D global profiles, matched to expectations and early science products from the MeerKAT \hi\ emission surveys LADUMA and MIGHTEE-HI. From these we considered their host galaxy properties with their asymmetry measures (lopsidedness, $A_L$, velocity offset, $A_{vo}$, and the channel-by-channel asymmetry, $\mathcal{A}_{min}$).

When comparing asymmetries to particular galaxy properties (gas mass, stellar mass, etc.) we always find large dispersions for each asymmetry statistic in each property bin. This is likely due to the large number of drivers of asymmetry, as well as the variations in profile asymmetry with line-of-sight. By keeping the ratio of `symmetric'/`asymmetric' galaxies the same for the total population, we arrive at new limits of $\mathrm{Lim}_{A_{\rm{vo}}}=0.06$ and $\mathrm{Lim}_{\mathcal{A}_{\rm{min}}}=0.19$ for this sample, which allowed us to explore the trends in greater detail across the different asymmetry measures. 


Examining the population as a whole, we find that the \hi\ mass has the strongest correlation with the profile asymmetries.  There are weaker correlations with stellar mass and the \hi\ gas fraction as well.  
In this analysis we do not see a significant relationship with the nearest neighbour distance for the global distribution. This could be due to confusion of \hi\ profiles due to flux from additional galaxies contained in the source cubes as well as merging galaxies not being classified as separate objects in the simulation catalogue. When separating the sample into previously merged and unmerged sub-samples, a difference in the symmetric fractions vs $D_{nn}$ can be seen for low $D_{nn}$ values where the merged sub-sample has a lower symmetric fraction (i.e. higher asymmetric fraction) than the unmerged sample possibly indicating a role of mergers in driving asymmetries.


When the populations are separated into mergers and non-mergers, the \hi\ mass dependence on asymmetries is more pronounced in the non-merger population.  When separated into low mass and high mass populations, a similar result is seen.  The low mass galaxies show a strong trend with the number of mergers, while the high mass galaxies do not show any trend.  This suggests that the relative effect of a merger is much greater on low mass galaxies than on high mass galaxies.  More interestingly, the decrease in $A_{L}$, $A_{vo}$, and $\mathcal{A}_{min}$ with the number of mergers in the low mass population indicates that repeated mergers in a short period tends to smooth out the gas distribution more than a single merger.

We compared our sample with the study of \cite{Watts2021} who considered the importance of deep \hi\ observations in discovering trends between symmetric and asymmetric populations in star-forming galaxies within the $\Delta f_{\rm gas}$--$\Delta$sSFR parameter space for ALFALFA and xGASS. We find agreement with \cite{Watts2021} in that the asymmetric, and especially strongly asymmetric, galaxies have lower \hi\ mass fractions than symmetric galaxies, and even at high \hi\ masses this weakly holds, where we see lower \hi\ gas fractions and specific star formation rates for asymmetric galaxies compared to symmetric galaxies. There is a large and obvious difference between the two populations when separating between high and low \hi\ masses for either symmetric or asymmetric galaxies. 

This work is merely the start of ASymba. There are multiple avenues to extend this study, including an extension to 2D and 3D asymmetry measures (Deg et al., in prep) and how these compare to the 1D case. It is already clear that asymmetries in different dimensions trace different aspects of galaxy growth requiring further investigation. The impact of other attributes, such as different classes of galaxy merger events, can be explored across all asymmetry measures described here and in higher dimensions. 


ASymba, alongside studies of \hi\ properties in \simba\ and other hydrodynamical simulations, will test our assumed galaxy models and enable direct comparison with the upcoming observational surveys that will greatly advance previous work, both in sample size and sensitivity, into the SKA era. 

\section*{Acknowledgements}

We thank the referee for useful suggestions that have helped improve the paper. We thank M. Verheijen for useful discussions.

MG acknowledges support from IDIA and was partially supported by the Australian Government through the Australian Research Council's Discovery Projects funding scheme (DP210102103). This work is based on the research supported in part by the National Research Foundation of South Africa (Grant number 129283). KS acknowledges support from the Natural Sciences and Engineering Research Council of Canada (NSERC).

The computing equipment to run \simba\ was funded by BEIS capital funding via STFC capital grants ST/P002293/1, ST/R002371/1 and ST/S002502/1, Durham University and STFC operations grant ST/R000832/1. DiRAC is part of the National e-Infrastructure.  We acknowledge the use of computing facilities of IDIA for part of this work. IDIA is a partnership of the Universities of Cape Town, of the Western Cape and of Pretoria. We acknowledge the use of the ilifu cloud computing facility - www.ilifu.ac.za, a partnership between the University of Cape Town, the University of the Western Cape, the University of Stellenbosch, Sol Plaatje University, the Cape Peninsula University of Technology and the South African Radio Astronomy Observatory. The ilifu facility is supported by contributions from the Inter-University Institute for Data Intensive Astronomy (IDIA - a partnership between the University of Cape Town, the University of Pretoria and the University of the Western Cape), the Computational Biology division at UCT and the Data Intensive Research Initiative of South Africa (DIRISA). This research made use of Astropy,\footnote{http://www.astropy.org} a community-developed core Python package for Astronomy \citep{astropy:2013, astropy:2018}.

\section*{Data availability}

The \simba\ simulation suite is publicly available at \url{http://simba.roe.ac.uk/}. The data underlying this article will be shared on reasonable request to the corresponding author.



\footnotesize{
  \bibliographystyle{mnras}
  \bibliography{bibliography}
}

\bsp	
\label{lastpage}
\end{document}